\documentclass[12pt,letter]{article}

\usepackage{graphicx, epsfig, color}
\usepackage{amsmath,amssymb,hyperref}
\usepackage{subfigure}

\def\eslt{\not\!\!{E_T}}
\def\to{\rightarrow}

\def\bi{\begin{itemize}}
\def\ei{\end{itemize}}
\def\te{\tilde e}

\def\tchi{\tilde\chi}

\def\tu{\tilde u}
\def\sps1ap{SPS1a$^\prime$}
\def\c1p{C1$^\prime$}

\def\tb{\tilde b}
\def\tf{\tilde f}

\def\tst{\tilde t}
\def\ttau{\tilde \tau}

\def\tg{\tilde g}
\def\tnu{\tilde\nu}

\def\alt{\lesssim}
\def\agt{\gtrsim}
\def\be{\begin{equation}}  
\def\ee{\end{equation}}  
\def\bea{\begin{eqnarray}}  
\def\eea{\end{eqnarray}}  
\def\beas{\begin{eqnarray*}}  
\def\eeas{\end{eqnarray*}}

\begin{document}
\begin{titlepage}
\begin{flushright}
OUHEP-231105
\end{flushright}

\vspace{0.5cm}
\begin{center}
{\Large \bf Natural anomaly-mediation from the landscape\\  
with implications for LHC SUSY searches
}\\ 
\vspace{1.0cm} \renewcommand{\thefootnote}{\fnsymbol{footnote}}
{\large Howard Baer$^1$\footnote[1]{Email: baer@nhn.ou.edu }, 
Vernon Barger$^2$\footnote[2]{Email: barger@pheno.wisc.edu },
Jessica Bolich$^1$\footnote[3]{Email: Jessica.R.Bolich-1@ou.edu}\\
Juhi Dutta$^1$\footnote[4]{Email: juhi.dutta@ou.edu} and
Dibyashree Sengupta$^3$\footnote[5]{Email: Dibyashree.Sengupta@lnf.infn.it}
}\\ 
\vspace{1.0cm} \renewcommand{\thefootnote}{\arabic{footnote}}
{\it 
$^1$Department of Physics and Astronomy,
University of Oklahoma, Norman, OK 73019, USA \\
}
{\it 
$^2$Department of Physics,
University of Wisconsin, Madison, WI 53706, USA \\
}
{\it
  $^3$ INFN, Laboratori Nazionali di Frascati,
Via E. Fermi 54, 00044 Frascati (RM), Italy}
\end{center}

\vspace{0.4cm}
\begin{abstract}
\noindent 
Supersymmetric models with the anomaly-mediated SUSY breaking (AMSB)
form for soft SUSY breaking terms arose in two different settings:
{\bf 1}) extra-dimensional models where SUSY breaking occurred in a
sequestered sector and 
{\bf 2}) $4-d$ models with dynamical SUSY breaking in a
hidden sector where scalars gain masses of order the gravitino mass $m_{3/2}$
but gaugino masses and trilinear soft terms assumed to be of the AMSB form.
Both models have run into serious conflicts with 1. LHC sparticle and Higgs
mass constraints, 2. constraints from wino-like WIMP dark matter searches
and 3. bounds from naturalness.
These conflicts may be avoided by introducing minor changes
to the underlying phenomenological models consisting of non-universal
bulk scalar Higgs masses and $A$-terms, providing a setting for
{\it natural anomaly-mediation} (nAMSB). In nAMSB, the wino is still expected
to be the lightest of the gauginos, but the higgsinos are expected to be
the lightest electroweakinos (EWinos) in accord with naturalness.
We examine what sort of spectra are expected to emerge when
nAMSB arises from a string landscape setting: while
model {\bf 2} can only be natural for a Higgs mass $m_h\alt 123$ GeV,
model {\bf 1} can accommodate naturalness along with $m_h\sim 125$ GeV
whilst still respecting LHC bounds on sparticle masses.
We explore the LHC phenomenology of nAMSB models where we find
that for higgsino pair production, typically larger dilepton
mass gaps arise from the soft dilepton-plus-jets signature than in models
with gaugino mass unificaton.
For wino pair production, the higher $m_{3/2}$ portion of nAMSB
parameter space is excluded by recent LHC bounds from
gaugino pair production searches.
We characterize the dominant LHC signatures arising from the remaining lower
$m_{3/2}\sim 90-200$ TeV range of parameter space which
should be fully testable at high-luminosity LHC via EWino pair
production searches.

\end{abstract}

\end{titlepage}

\section{Introduction}
\label{sec:intro}

Supersymmetric models\cite{Chung:2003fi} based on  anomaly-mediated SUSY breaking (AMSB)
arose from two different set-ups.

\subsection{AMSB0: (GLMR)}

The second, by Giudice {\it et al.}\cite{Giudice:1998xp} (GLMR),
which we label as $AMSB0$, was motivated by
4-dimensional models where SUSY is broken dynamically in the hidden
sector\cite{Witten:1981nf}, and where
SUSY breaking was communicated to the visible sector via gravity.
The motivation here was that the SUSY breaking scale $m_{hidden}$
might be generated non-perturbatively via gaugino condensation and would then
be exponentially-suppressed relative to the Planck scale
via dimensional transmutation\cite{Dine:2010cv}:
$m_{hidden}\sim e^{-8\pi^2/g^2} m_P$, where $m_P$ is the reduced Planck scale.
This would not only stabilize the weak scale (via SUSY), but also explain
its exponential suppression from the Planck scale:
$m_{weak}\sim m_{soft}\sim m_{hidden}^2/m_P$, where $m_{hidden}\sim 10^{11}$ GeV.
Now in gravity mediation, gaugino masses arise via
\be
\int d^2\theta f_{AB}(\frac{S}{m_P})W^A_\alpha W^{B\alpha}
\ee
with $f_{AB}$ the gauge kinetic function 
depending on hidden sector fields $S$ and where the $F$-term of $S$
acquires a SUSY breaking value $F_S$;
the gaugino masses arise as $m_\lambda\sim (F_S/m_P)\sim m_{3/2}\sim m_{soft}\sim m_{weak}$.
However, if no hidden sector singlets are available as in
(most) DSB models\cite{Affleck:1984xz}\footnote{For an exception, see {\it e.g.} \cite{Nelson:1995hf}.}, then
the gaugino masses are expected instead at the keV scale, which would be
experimentally excluded.
However, Ref. \cite{Giudice:1998xp} found that the one-loop renormalization
of the visible sector gauge couplings is given by\cite{LopesCardoso:1993sq}
\be
\frac{1}{4}\int d^2\theta\left(1-\frac{g^2b_0}{16\pi^2}\log
\frac{\Lambda^2}{\Box }\right) W^\alpha W_\alpha +h.c.
\ee
where $b_0$ is the coefficient of the relevant gauge group beta-function
and $\Box$ is the d'Alembertian operator.
This leads to SUSY breaking gaugino masses via replacement of the UV cutoff
$\Lambda$ by the spurion superfield $\Lambda\exp (m_{3/2} \theta^2)$
leading to (loop-suppressed) gaugino masses\cite{Bagger:1999rd}\footnote{
  The AMSB contributions to gaugino masses were already presaged by
  Ref's \cite{Dixon:1990pc}, \cite{Dine:1992yw} and \cite{Kaplunovsky:1993rd,Kaplunovsky:1994fg}.}
\be
m_\lambda =-\frac{g^2b_0}{16\pi^2}m_{3/2} .
\ee
For $m_{3/2}\sim 100$ TeV, then $m_\lambda\sim 1$ TeV as required
to gain $m_{weak}\sim m_{W,Z,h}\sim  100$ GeV.
Similarly, the trilinear soft ($A$)-terms are not allowed at tree level if no
singlets are available for a
\be
\int d^2\theta \frac{S}{m_P}\phi_i\phi_j\phi_k
\ee
coupling (where the $\phi_i$ are generic visible sector superfields).
The $A$-terms can also arise at one-loop level in AMSB and are proportional to
derivatives of the anomalous dimensions.
Scalar masses on the other hand arise from
\be
\int d^2\theta d^2\bar{\theta}\frac{S^\dagger S}{m_P^2}\phi^\dagger\phi
\ee
and are not protected by symmetries and so can be much larger,
$m_\phi^2\sim m_{3/2}^2$, and can gain their gravity-mediated form.
This form of scalar mass generation suffers the usual SUSY flavor problem
that is endemic to gravity-mediation.

The AMSB0 model thus yields a hierarchy of soft terms
$m_\phi\gg m_\lambda\sim A$ as noted by Wells\cite{Wells:2003tf}
in what he dubbed PeV-SUSY\cite{Wells:2004di}.
This model also motivated realizations of
split\cite{Arkani-Hamed:2004ymt,Arkani-Hamed:2004zhs}-
and minisplit\cite{Arvanitaki:2012ps} SUSY models.
These later models eschew the notion of naturalness in hopes of a landscape
solution to the naturalness problem, thus allowing for scalar masses
in the range of 100-1000 TeV (for minisplit) and ranging up
to $m_{\phi}\sim 10^9$ TeV for split SUSY.
Split SUSY predicts a light Higgs mass $m_h\sim 130-160$
GeV\cite{Giudice:2011cg}.
The discovery of a SM-like Higgs boson with mass $m_h\sim 125$ GeV motivated
a retreat to scalars in the range of minisplit models which allow for
$m_h\sim 125$ GeV along with small $A$-terms.
A value of $m_h\simeq 125$ GeV can also be realized by TeV-scale top squarks
but with near maximal stop mixing from large $A$-terms\cite{Carena:2002es,Baer:2011ab}.

Since scalar masses arise as in gravity-mediation, this $AMSB0$ model
may still be plagued by flavor problems, although these may be softened
by the rather large values of scalar masses which are expected:
a (partial) decoupling solution to the SUSY flavor problem\cite{Dine:1993np}.
It also gave rise to unique phenomenological signatures\cite{Feng:1999fu}
since in AMSB  the {\it wino} rather than the bino was expected
to be the lightest SUSY particle (LSP).

\subsection{AMSB (RS)}

Alternatively, in the Randall-Sundrum AMSB model\cite{Randall:1998uk}
($AMSB$),
it was posited that SUSY breaking arose in a
hidden sector sequestered from the visible sector in extra-dimensional
spacetime. In such a set-up, the leading contribution to
{\it all} soft SUSY breaking terms was from the superconformal anomaly,
and suppressed by a loop factor from the gravitino mass $m_{3/2}$. 
In this form of AMSB, a common value of scalar masses was expected thus
avoiding the SUSY flavor problem which seems endemic to models of
gravity-mediation.
Also, since $m_{soft}\ll m_{3/2}$, the cosmological gravitino
problem could be avoided since in the early universe
thermally-produced gravitinos could decay before the
onset of BBN\cite{Kawasaki:2008qe}.
In both cases of $AMSB$ and $AMSB0$, the thermally underproduced
wino-like WIMPs could have their relic abundance non-thermally enhanced by
either gravitino\cite{Pradler:2006qh} or moduli-field decays\cite{Moroi:1999zb,Bae:2022okh}.
A drawback in the case of $AMSB$ was that soft slepton masses
were derived to be {\it tachyonic} thus leading to charge-breaking vacua
in the scalar potential.
Some extra contributions to scalar masses arising from fields propagating
in the bulk of spacetime could be postulated to avoid this
problem\cite{Randall:1998uk}.

\subsection{Further deliberations on AMSB}

Some further notable theoretical explorations of AMSB soft terms
include Gaillard {\it et al.}\cite{Gaillard:2000fk}
  where AMSB soft terms arose as quantum corrections under Pauli-Villars
  regularization of supergravity.
  In Anisimov {\it et al.}\cite{Anisimov:2001zz,Anisimov:2002az},
  brane world SUSY breaking (as in RS model) was examined, and it was found
  to be insufficient to guarantee the needed sequestering between hidden
  and observable sectors to generate dominant AMSB soft terms and
  flavor-conserving scalar masses.
  In Ref. \cite{Luty:2005sn}, Luty presents pedagogical lectures on
  SUSY breaking leading up to and including AMSB.
  In Ref. \cite{Jones:2006re}, the connection of AMSB to dimensional
  transmutation is examined
  as a solution to the tachyonic slepton problem.
  In Ref. \cite{Dine:2007me}, Dine and Seiberg (DS) clarify the derivation of AMSB soft terms and relate them to the gaugino counterterm.
  In Ref. \cite{deAlwis:2008aq}, de Alwis presents the derivation of
  AMSB soft terms and emphasizes their origin in work by
  Kaplunovsky and Louis\cite{Kaplunovsky:1994fg} and DS,
  and shows there may be additional soft term contributions.
  This inspires his later development of the gaugino AMSB
  model\cite{Baer:2010uy}.
 In Ref. \cite{Jung:2009dg}, a clarifying derivation of AMSB soft terms is presented.
  In Ref. \cite{Sanford:2010hc}, Sanford and Shirman
  develop an arbitrary conformal compensator formalism which allows
  extrapolation between RS and DS derivations.
 In Ref. \cite{Conlon:2010qy}, anomaly mediation from IIB string theories
  is examined.
  In Ref. \cite{DEramo:2012vvz}, the AMSB connection to gravitino mediation vs. K\"ahler mediation is examined.
  This work is extended to scalar masses in Ref. \cite{DEramo:2013dzi}. 
  In Ref. \cite{Dine:2013nka},
  Dine and Draper examine anomaly mediation in local effective theories.
  In Ref. \cite{deAlwis:2012gr}, de Alwis examines the interplay of AMSB
  with spontaneous SUSY breaking.
  In Ref. \cite{Harigaya:2014sfa}, the connection between AMSB
  gaugino masses and the path integral measure is examined.

  An alternative route to models with AMSB soft terms was developed
  by Luty and Sundrum\cite{Luty:2001zv},
  in models with strong hidden sector conformal dynamics.
  In these $4-d$ models, strong hidden sector conformal dynamics leads
  to a suppression, or sequestering, of usual soft terms due to higher
  dimensional operators which mix the hidden and visible sectors.
  The suppression of gravity-mediated soft terms occurs between the
  messenger scale (taken here to be $m_P$) and some intermediate scale
  $m_{int}$ where conformal symmetry becomes broken. In such a case,
  the loop-suppressed AMSB soft terms may become dominant. In
  Ref. \cite{Dine:2004dv}, the conformal suppression acts upon scalar masses
  and the $B\mu$ term, but in Ref's \cite{Ibe:2005pj,Murayama:2007ge}
  it is emphasized that conformal sequestering may also act on the
  gaugino sector.

\subsection{Status of minimal phenomenological AMSB model (mAMSB)}

A minimal phenomenological AMSB model (mAMSB) was proposed in
Ref's \cite{Gherghetta:1999sw} and \cite{Feng:1999hg} with parameter space
\be
m_0,\ m_{3/2},\ \tan\beta,\ sign(\mu )\ \ \ (mAMSB)
\ee
where $m_0$ was an added universal bulk scalar mass and the
gravitino mass $m_{3/2}$ set the scale for the AMSB soft terms
$m_{AMSB}\sim c(g^2/16\pi^2)m_{3/2}$ with $c$ a calculable constant of
order unity and $g$ is a gauge group coupling constant.
The bulk scalar mass is generic to the $AMSB0$ set-up and
phenomenologically required to gain positive slepton squared
masses in $AMSB$.
Various studies for mAMSB at LHC appeared in Ref's
\cite{Paige:1999ui,Baer:2000bs,Barr:2002ex,Allanach:2011qr,Bagnaschi:2016xfg}.

At present, both these set-ups within the mAMSB model seem phenomenologically
disfavored and perhaps even ruled out.
The first problem is that in mAMSB the
SUSY conserving $\mu$ parameter is typically fine-tuned to large values
compared to the measured value of the weak scale $m_{weak}\sim m_{W,Z,h}\sim 100$ GeV, thus violating\cite{Baer:2014ica} even the most conservative measure of
naturalness $\Delta_{EW}$\cite{Baer:2012up,Baer:2012cf},
where $\Delta_{EW}$ is defined as the largest value on the right-hand-side (RHS)
of the scalar potential minimization condition
\be
m_Z^2/2=\frac{m_{H_d}^2+\Sigma_d^d-(m_{H_u}^2+\Sigma_u^u)\tan^2\beta}{\tan^2\beta -1}-\mu^2
\label{eq:mzs}
\ee
divided by $m_Z^2/2$.
The second problem is that the small values of
mAMSB $A$-terms typically lead to too small a value of $m_h\ll 125$ GeV
unless third-generation soft scalar masses lie in the 10-100 TeV
range\cite{Arbey:2011ab,Baer:2012uya} thus also violating
naturalness\cite{Baer:2014ica} via the radiative
corrections $\Sigma_u^u(\tst_{1,2})$,
leading again to a large value for the
electroweak fine-tuning measure $\Delta_{EW}$.
A third problem is that wino-only dark matter\cite{Moroi:1999zb}
now seems excluded by a combination of direct and indirect WIMP
search experiments\cite{Cohen:2013ama,Fan:2013faa,Baer:2016ucr}.
This latter exclusion may be circumvented in cases of mixed axion-wino
dark matter (two DM particles) wherein the relic wino abundance forms only
a small portion of the total DM abundance\cite{Bae:2015rra}.
This latter scenario posits a PQ axion which also solves the finetuning
problem of the $\theta$ parameter in the QCD sector.

\subsection{Natural anomaly-mediated SUSY breaking (nAMSB)}

In Ref. \cite{Baer:2018hwa}, two minor changes to the mAMSB model were
suggested to circumvent its undesirable phenomenological properties.\footnote{
  These ``changes'' were actually suggested as default parameters
  in the original RS paper Ref. \cite{Randall:1998uk}.}
First, separate bulk masses for $m_{H_u}\ne m_{H_d}\ne m_0$ were applied to scalar
masses which then allowed for a small $\mu$ parameter in accord with
naturalness (a unified mass $m_0$ just for matter scalars is highly motivated
by the fact that the matter superfields are unified within the
16-dimensional spinor rep of $SO(10)$).
Second, bulk contributions to trilinear soft terms $A_0$
were advocated which then allowed for large stop mixing which in turn
uplifts the Higgs mass $m_h\to \sim 125$ GeV\cite{Slavich:2020zjv}
without requiring the stop sector to lie in the unnatural multi-TeV range
or beyond.
These two adjustments allowed for EW naturalness and for $m_h\sim 125$ GeV.
Models with these attributes were denoted as {\it natural} AMSB (nAMSB):
\be
m_0(i),\ m_{H_u},\ m_{H_d},\ m_{3/2},\ A_0,\ \tan\beta \ \ \ (nAMSB^\prime )
\label{eq:nAMSBp}
\ee
where we also allow for possible non-universal bulk contributions to the
different generations $i=1-3$.
It is convenient to then trade the high scale parameters $m_{H_u}^2$ and
$m_{H_d}^2$ for weak scale parameters $\mu$ and $m_A$ using the
scalar potential minimization conditions\cite{Ellis:2002wv,Baer:2005bu}.

Like mAMSB, the nAMSB model has winos as the lightest of the gauginos.
Unlike mAMSB, the nAMSB model (usually) has
{\it higgsinos as the lightest EWinos}, in accord with naturalness.
By requiring the natural axionic solution to the strong CP problem,
one then expects mixed axion plus higgsino-like WIMP dark matter\cite{Baer:2011hx,Bae:2013hma},
and one can circumvent the constraints on wino-only dark matter\cite{Baer:2016ucr}.

\subsection{nAMSB from the landscape}

The advent of the string theory landscape\cite{Bousso:2000xa,Susskind:2003kw}
led to some major changes in SUSY models with AMSB soft terms.
First, it was found that flux compactification\cite{Douglas:2006es}
of type IIB string models on Calabi-Yao orientifolds led to
enormous numbers of string vacuum states ($10^{500}$ is a prominently
quoted number\cite{Douglas:2003um,Ashok:2003gk} although much larger numbers have also been found in
$F$-theory compactifications\cite{Taylor:2015xtz}).
Such large numbers of vacuum possibilities allow for
Weinberg's\cite{Weinberg:1987dv} anthropic solution to the cosmological
constant (CC) problem and ``explains'' the finetuning of $\Lambda_{CC}$
to a part in $10^{120}$.
Then, if the CC is finetuned by anthropics, might one also
allow for the little hierarchy $m_{weak}\ll m_{soft}$ to also be finetuned?
In split SUSY, electroweak naturalness is eschewed while WIMP dark matter and
gauge coupling unification are retained\cite{Wells:2003tf,Arkani-Hamed:2004ymt,Giudice:2004tc}.
A possible model framework for split SUSY would then be charged SUSY
breaking\cite{Arkani-Hamed:2004zhs,Wells:2004di}, wherein
tree level gaugino masses (and $A$-terms) are forbidden by some symmetry
(perhaps $R$-symmetry?)
while scalar masses are allowed as heavy as one likes.
Values of $m_{scalar}\sim 10^9$ GeV were entertained, leading to a signature
of long-lived gluinos.
The heavy scalars also allowed for a decoupling solution to the SUSY
flavor and CP problems\cite{Dine:1993np}.
In split SUSY, one expects light Higgs masses in the
$m_h\sim 130-160$ GeV range\cite{Giudice:2011cg,Bagnaschi:2014rsa},
in contrast to the 2012 Higgs discovery with $m_h\simeq 125$ GeV.
To accommodate the measured Higgs mass,
scalar masses were dialed down to the $10^3$ TeV range.
These {\it minisplit} models\cite{Arvanitaki:2012ps,Arkani-Hamed:2012fhg}
then allowed for $m_h\sim 125$ GeV while still potentially allowing for a
decoupling solution to the SUSY flavor and CP problems.

However, only recently has the occurrence frequency of highly finetuned
SUSY models been examined in an actual landscape context.
In Ref. \cite{Baer:2022wxe}, a toy model of
the landscape was developed, and it was shown that EW {\it natural} models
should be more likely than finetuned models to emerge from a generic
landscape construction. In retrospect, the reason is rather simple.
In Agrawal {\it et al.}\cite{Agrawal:1997gf} (ABDS),
it was found that within a multiverse
wherein each pocket universe would have a different value for its weak scale,
then only complex nuclei, and hence complex atoms
(which seem necessary for life as we know it) would arise if the
pocket universe value of the weak scale were within a factor of a few
of its measured value in our universe (OU):
$0.5 m_{weak}^{OU}\alt m_{weak}^{PU}\alt 5 m_{weak}^{OU}$.
We call this range of $m_{weak}^{PU}$ the ABDS window.
Now in models where all contributions to the weak scale
(to the RHS of Eq. \ref{eq:mzs}) are
natural (in that they lie within the ABDS window), then the remaining parameter
selection (typically either $\mu (weak)$ or $m_{H_u}(weak)$) will also have a
wide range of possibilities, all lying within the ABDS window, to gain an
ultimate value for $m_{weak}$ within the ABDS window. On the other hand,
if any contribution to $m_{weak}$ is far beyond $m_{weak}$, then finetuning is
needed and only a tiny portion of parameter space will lead to
$m_{weak}\sim 100$ GeV.
This scheme was then used in Ref. \cite{Baer:2022dfc} to compute
relative probabilities $P_\mu$ for different natural and finetuned SUSY models
(and the SM) to emerge from the landscape.
For instance, from Ref. \cite{Baer:2022dfc}, it was found that for a
radiative natural SUSY model,
where all contributions to the weak scale lie within the ABDS window,
a relative probability $P_\mu\sim 1.4$ was computed whilst the SM,
valid up to the reduced Planck mass $m_P$, had $P_\mu\sim 10^{-26}$.
Also, split SUSY-- with scalar masses at $10^6$ TeV-- had $P_\mu\sim 10^{-11}$.
Other models such as CMSSM\cite{Kane:1993td}, PeV-SUSY\cite{Wells:2004di},
spread SUSY\cite{Hall:2011jd}, minisplit\cite{Arvanitaki:2012ps},
high-scale SUSY\cite{Barger:2005qy} and $G_2MSSM$\cite{Acharya:2008zi}
were also examined and found to have tiny values of $P_\mu$.
Thus, while the emergence of EW fine-tuned models is logically
possible from the landscape, their liklihood is highly suppressed compared to
natural models: natural SUSY models are much more plausible as a
low energy effective field theory (LE-EFT) realization of the string landscape.

With the above considerations in mind, in this paper we first wish to explore
in Sec. \ref{sec:scan1} the expectations for Higgs boson and
sparticle masses from the nAMSB model with sequestered sector SUSY breaking--
as might be expected from SUSY brane-world models, and as characterized
by the presence of bulk $A$-terms ($A_0\ne 0$ but also including AMSB
$A$-terms).
The nAMSB0 model with $A_0=0$ has been shown in Fig. 2 of Ref. \cite{Baer:2018hwa} to allow for naturalness ($\Delta_{EW}\alt 30$) but only if $m_h\alt 123$ GeV.
With the string landscape in mind,
we expect the various bulk soft terms and $m_{3/2}$ to be distributed
as a power-law draw to large values in the multiverse as suggested by
Douglas\cite{Douglas:2004qg}.
By combining the draw to large soft terms with the
requirement of a weak scale within the ABDS window, then the
putative distribution of Higgs and sparticle masses from the landscape
may be derived in the context of those string models which reduce to
a nAMSB low energy effective theory.
Generically, under charged SUSY breaking with gravity-mediated scalar masses,
we expect non-universality within different GUT multiplets and different
generations, so we adopt independent masses $m_0(i)$, ($i=1-3$ a
generation index),
along with $m_{H_u}\ne m_{H_d}$. Motivated by the fact that all members of
each generation fill out a complete 16-d spinor of $SO(10)$, we maintain
universality within each generation
(as emphasized by Nilles {\it et al.}\cite{Nilles:2014owa}).
One issue is that the bulk trilinear soft terms $A_0$ are expected to
be forbidden under charged SUSY breaking\cite{Wells:2004di}.
These results show the difficulty of deriving $m_h\sim 125$ GeV in such
models without bulk $A$-terms.
Thus, our ultimate parameter space is
\be
m_0(i),\ m_{3/2},\ A_0,\ \mu,\ m_A,\ \tan\beta\ \ \ (nAMSB) .
\label{eq:nAMSB}
\ee
We then restrict ourselves to a set of string landscape vacua with the MSSM
as the low energy EFT, but where gauginos gain AMSB masses, but the remaining
soft terms scan within the multiverse and include bulk terms. 
(While soft terms are expected to be correlated within our universe,
they may scan within the multiverse\cite{Baer:2020vad}.)


With this setup in mind, the remainder of this paper is organized as follows.
In Sec. \ref{sec:scan1}, we assume a simple $n=+1$ power-law draw on soft terms in the landscape, and plot out probability functions for the various expected
Higgs and sparticle masses for $nAMSB$ with $A_0\ne 0$.
These models can be natural whilst also respecting $m_h\sim 125$ GeV.
In Sec. \ref{sec:BM}, we present several  $AMSB$ benchmark points
and  model lines.
In Sec. \ref{sec:prod}, we present sparticle production cross sections
expected from nAMSB along our given model line.
Here, we find that typically higgsino and wino pair production is dominant over
the entire range of $m_{3/2}$ values.
In Sec. \ref{sec:decay}, we discuss the wino decays in nAMSB
for the dominantly produced sparticles.
In Sec's \ref{sec:lhc_excluded},
we discuss the main signal channels expected for LHC searches for nAMSB.
Given the AMSB weak scale gaugino mass ratio $M_1:M_2:M_3\sim 3:1:8$,
it is possible that strong new limits on gaugino
pair production from LHC could exclude $nAMSB$ up to and
perhaps even beyond its naturalness limit. However, there remains a
low mass window with $m(wino)\agt m(higgsino)$ which is still allowed due
to the semi-compressed spectrum of the EWinos.
After implementing present LHC constraints on nAMSB parameter space,
in Sec. \ref{sec:lhc} we discuss the most favorable avenues for
future SUSY searches within the nAMSB framework: via higgsino and wino pair
production.
Our summary and conclusions follow in Sec. \ref{sec:conclude}.

\section{Sparticle and Higgs masses in $nAMSB$ from the landscape}
\label{sec:scan1}

Here, we scan over parameters with a landscape-motivated $m_{soft}^1$
(linear) draw to large soft terms\cite{Douglas:2004qg,Baer:2017uvn}:
\bi
\item $m_{3/2}:\ 80-400\ {\rm TeV}$,
\item $m_0(1,2):\ 1-20\ {\rm TeV}$,
\item $m_0(3):\ 1-10\ {\rm TeV}$,
\item $A_0:\ 0- \pm 20\ {\rm TeV}$,
\item $m_A:\ 0.25-10\ {\rm TeV}$,
  \item $\tan\beta :\ 3-60$ (flat scan).
\ei
In accord with naturalness, we fix $\mu =250$ GeV. In lieu of requiring
the pocket-universe value of $m_Z^{PU}$ to lie within the ABDS window,
we instead invoke $\Delta_{EW}<30$ to avoid finetuning from terms beyond
the ABDS window: the finetuned solutions are much more rare compared to
non-finetuned (natural) solutions because in the finetuned case the
scan parameter space rapidly shrinks to a tiny interval\cite{Baer:2022wxe,Baer:2022dfc}. For the present case, we restrict the landscape to those vacuum
solutions which lead to the nAMSB model as the low energy effective field
theory, but where the contributions to the soft breaking terms scan over
this restricted portion of the multiverse.

\begin{figure}[tbp]
\begin{center}
    \includegraphics[height=0.3\textheight]{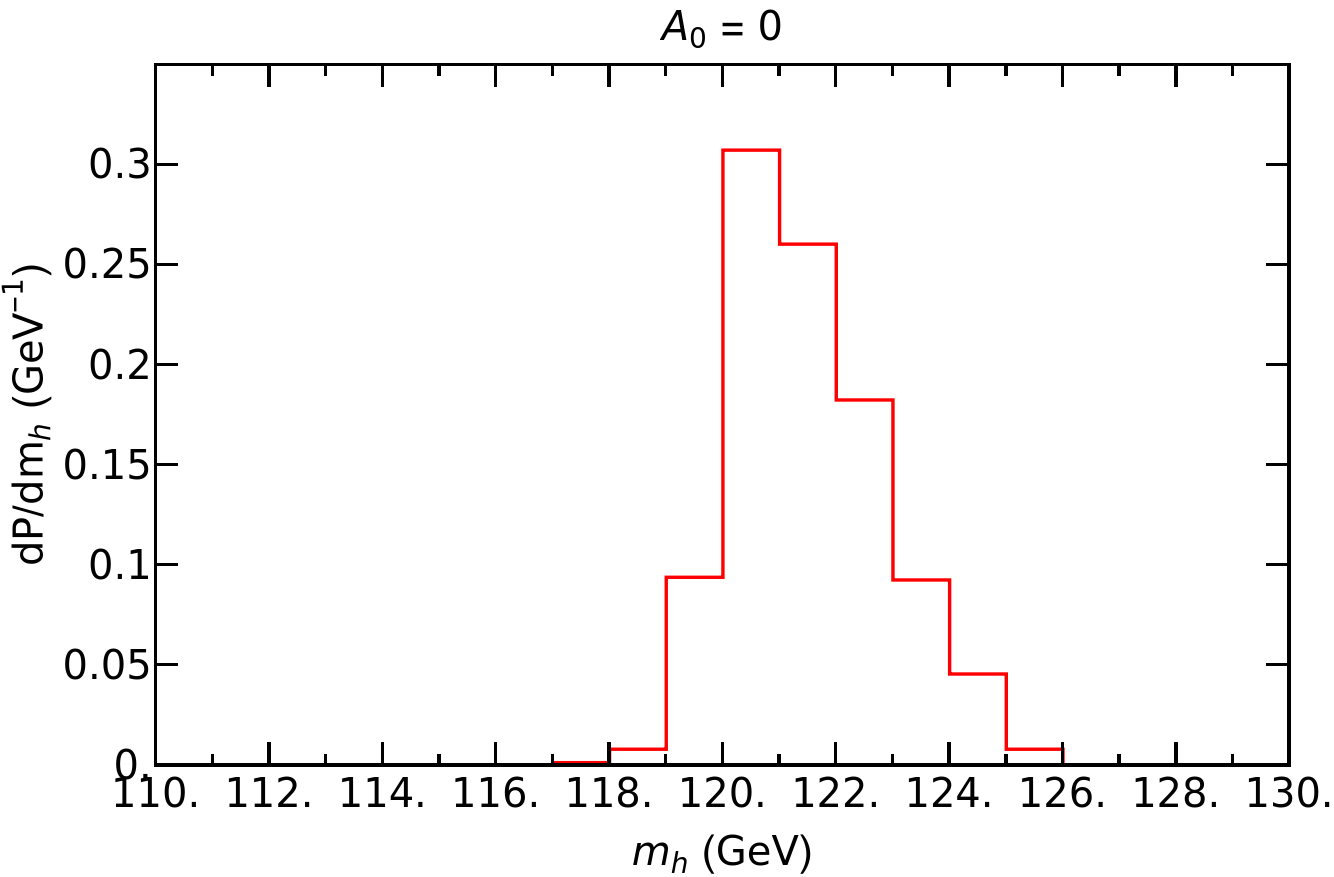}
  \caption{Plot of $dP/dm_h$ 
from an $n=1$ landscape scan in the nAMSB0 model where $A_0=0$.
\label{fig:scan0}}
\end{center}
\end{figure}

Our first results are shown in Fig. \ref{fig:scan0} for the
nAMSB0 model where $A_0$ is fixed at zero. In this case, we see that the
probability distribution peaks at $m_h\sim 120$ GeV and falls sharply
with increasing $m_h$. While some probability still exists for
$m_h\sim 125$ GeV, we henceforth move beyond nAMSB0 to the
nAMSB model with $A_0\ne 0$ where prospects for generating a
Higgs mass $m_h$ in accord with LHC data are much better.

\begin{figure}[tbp]
\begin{center}
  \includegraphics[height=0.2\textheight]{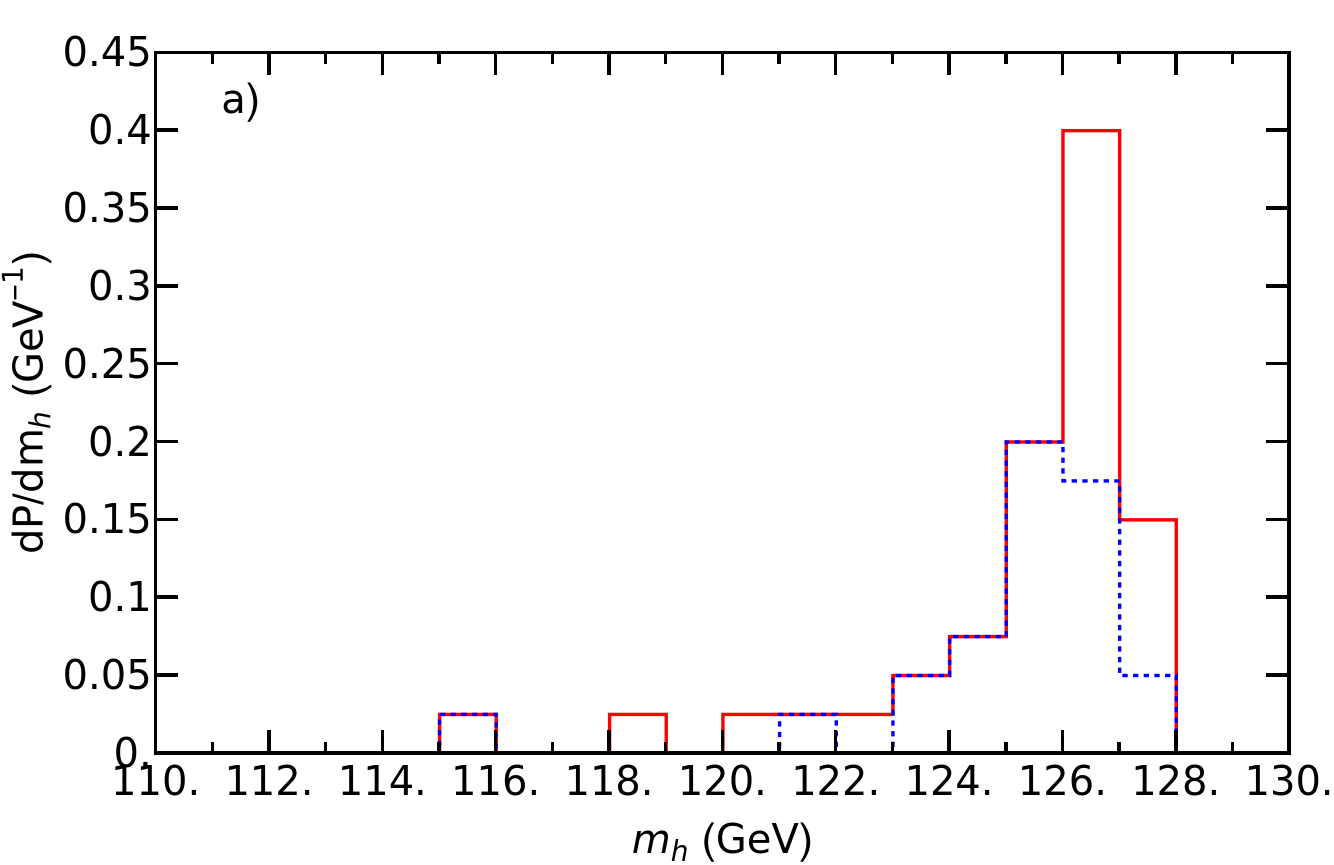}
  \includegraphics[height=0.2\textheight]{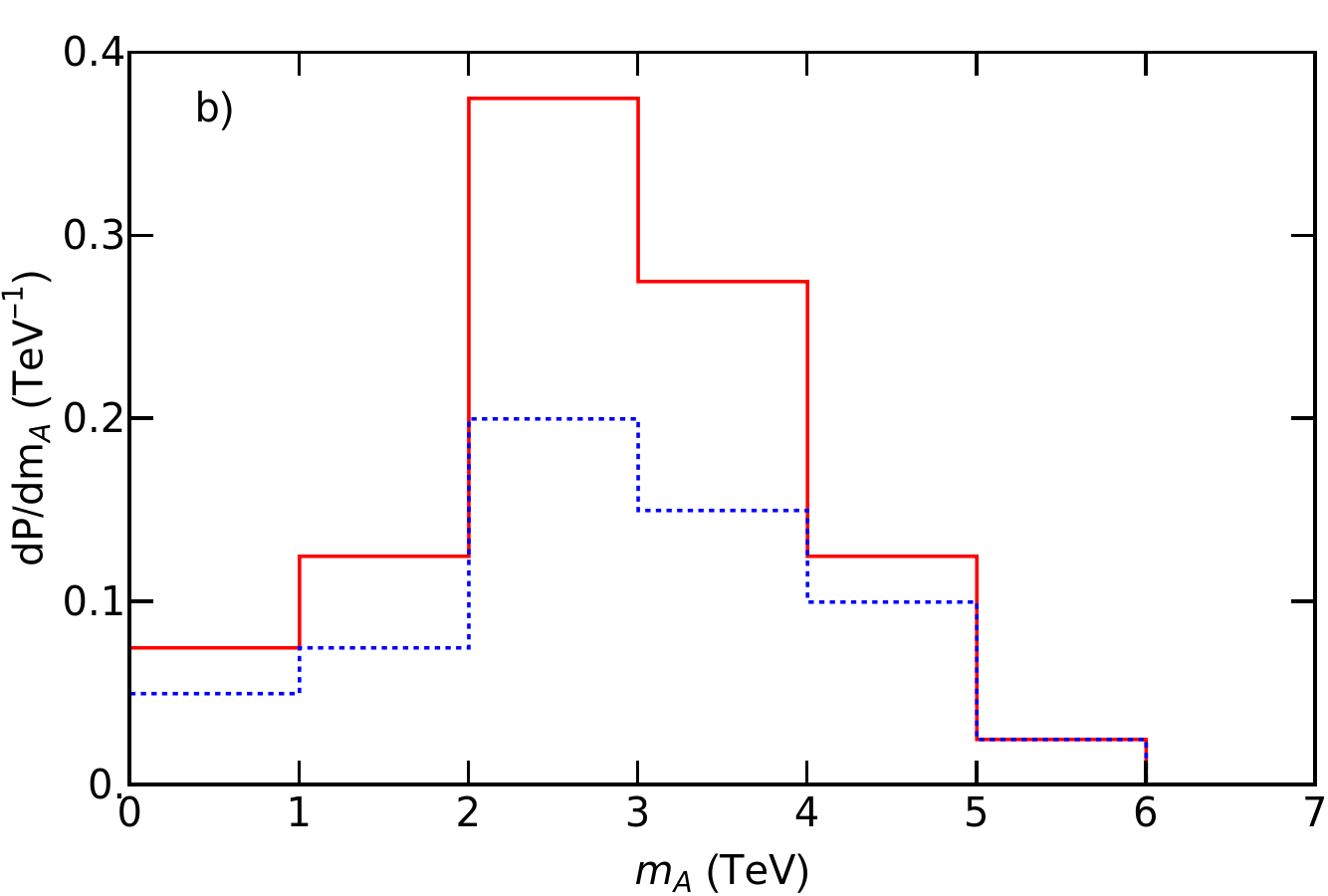}
  \caption{Plot of {\it a}) $dP/dm_h$ and {\it b}) $dP/dm_A$,
from an $n=1$ landscape scan in the nAMSB model.
The red histogram shows the full probability distribution while the
blue-dashed histogram shows the remaining distribution after
LHC sparticle mass limits are imposed.
\label{fig:scan1Higgs}}
\end{center}
\end{figure}

Our first results for nAMSB are shown in Fig. \ref{fig:scan1Higgs}.
In frame {\it a}), we show the differential probability distribution
$dP/dm_h$ vs. $m_h$, where $P$ is the probability normalized to
unity.
The red histogram shows the full probability distribution while the blue-dashed
histogram shows the same distribution after LHC sparticle mass limits
(discussed in Sec. \ref{sec:lhc_excluded}) are imposed.
We see that $dP/dm_h$ has only small values for $m_h\alt 123$ GeV,
but then peaks sharply in the range $m_h\sim 125-127$ GeV. This is in accord
with similar results in models with unified gaugino masses\cite{Baer:2017uvn}
or mirage-mediated gaugino masses\cite{Baer:2019tee}: basically, the soft terms
$m_0(1,2)$, $m_0(3)$, $A_0$, $m_A$ and $m_{3/2}$ are selected to be as large
as possible subject to the condition that the derived value of
$m_Z^{PU}$ lies within the ABDS window. This pulls the stop soft terms
$m_0(3)$ large into the $\sim 5$ TeV range (but not too large) and also
the bulk term $A_0$ to large-- nearly maximal-- mixing values, but not so large
as to lead to CCB minima of the scalar potential (CCB or no-EWSB minima
must be vetoed as not leading to a livable universe as we know it).
These conditions pull $m_h$ up to the vicinity of $\sim 125$ GeV.
We also show in frame {\it b}) the distribution in pseudoscalar Higgs mass
$m_A$, where $m_A$ contributes directly to the weak scale through
Eq. \ref{eq:mzs} since for $m_{H_d}\gg m_Z$, then $m_A\simeq m_{H_d}$
(and $m_H\sim m_{H^\pm}\sim m_A$). Here, we see that $m_A$ reaches peak
probability around $\sim 2.5$ TeV, somewhat beyond the reach of
HL-LHC\cite{Baer:2022qqr}. Maximally, $m_A$ can extend up to $\sim 6$ TeV
before overcontributing to the weak scale. 

\begin{figure}[tbp]
\begin{center}
  \includegraphics[height=0.2\textheight]{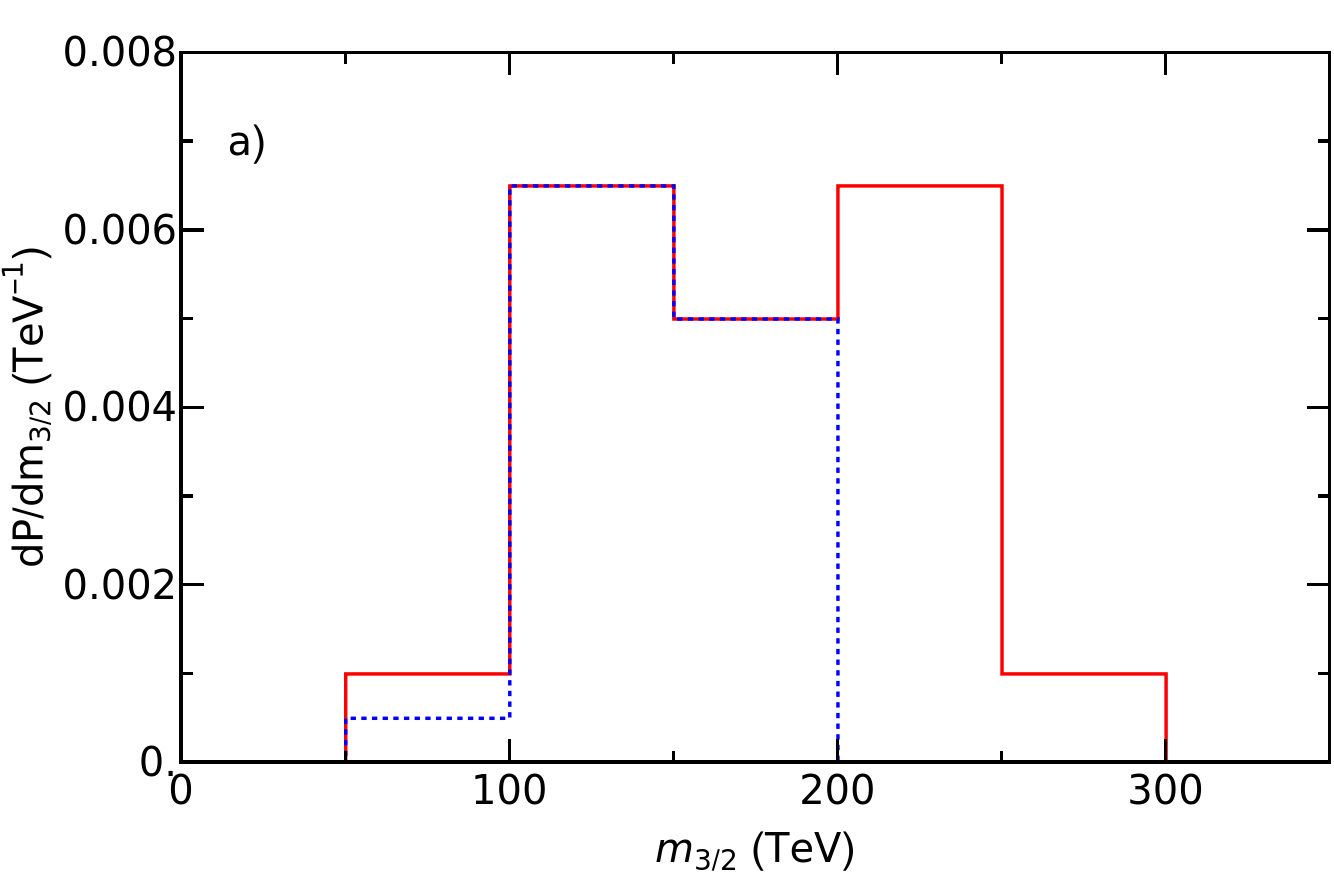}
  \includegraphics[height=0.2\textheight]{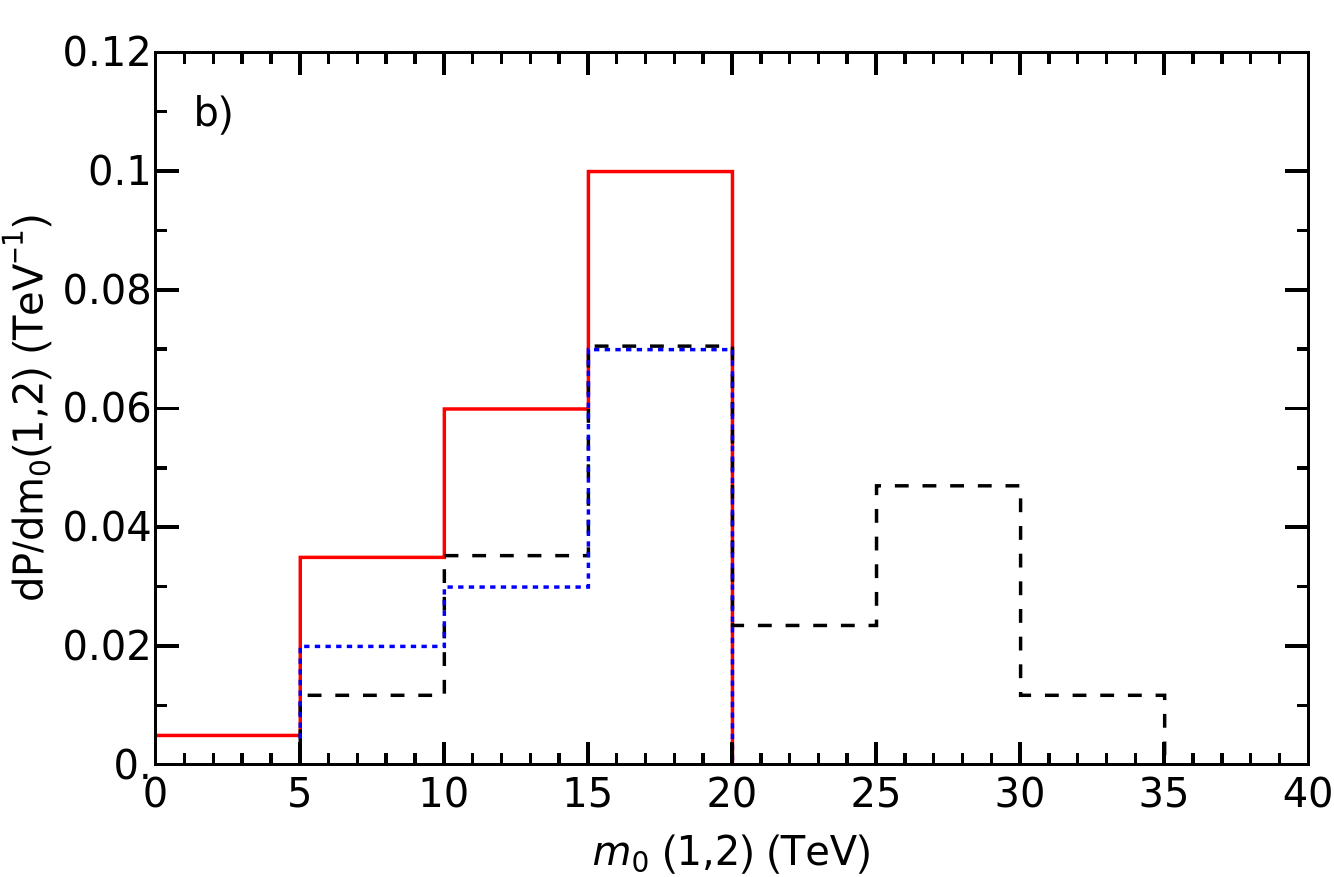}\\
  \includegraphics[height=0.2\textheight]{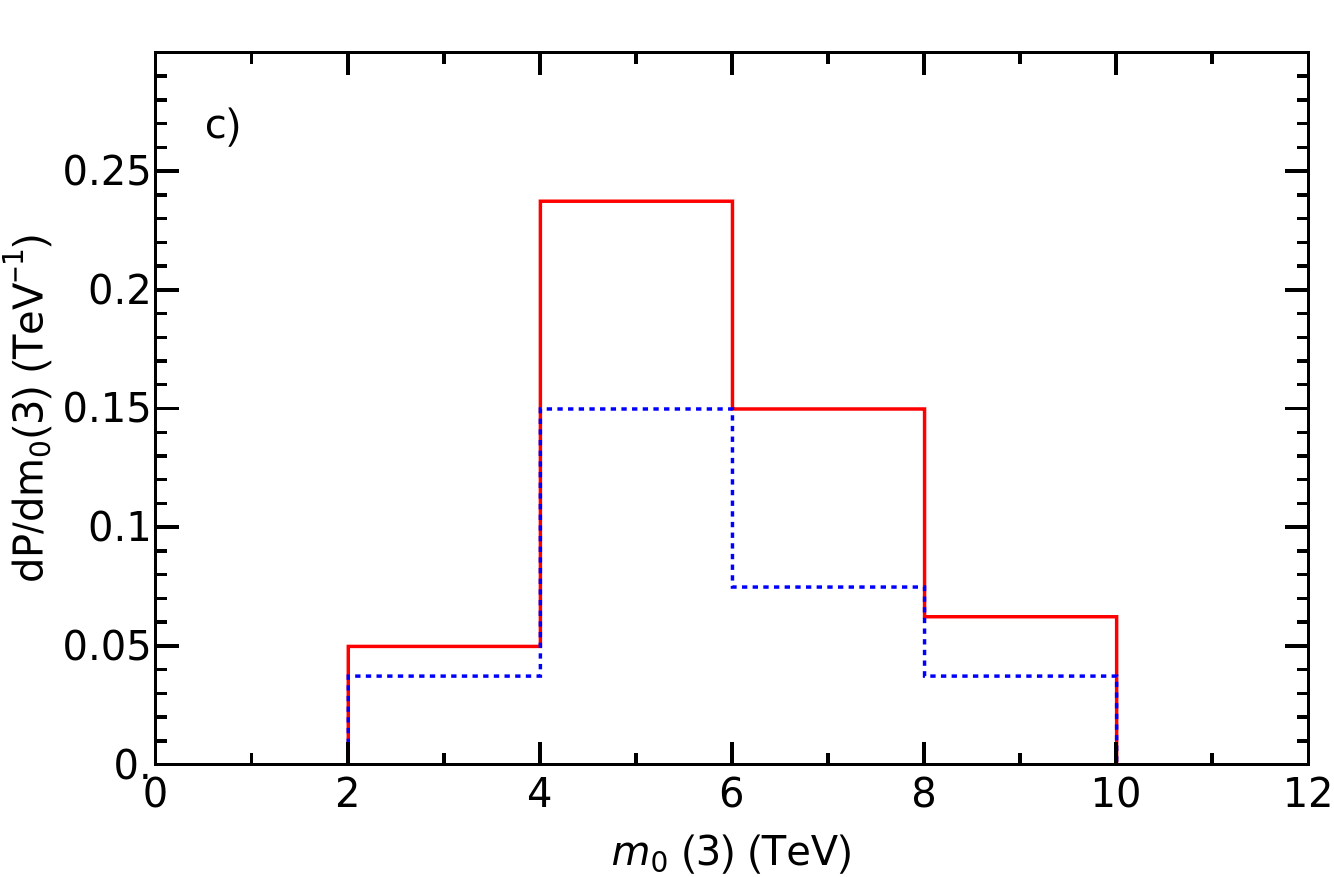}
  \includegraphics[height=0.2\textheight]{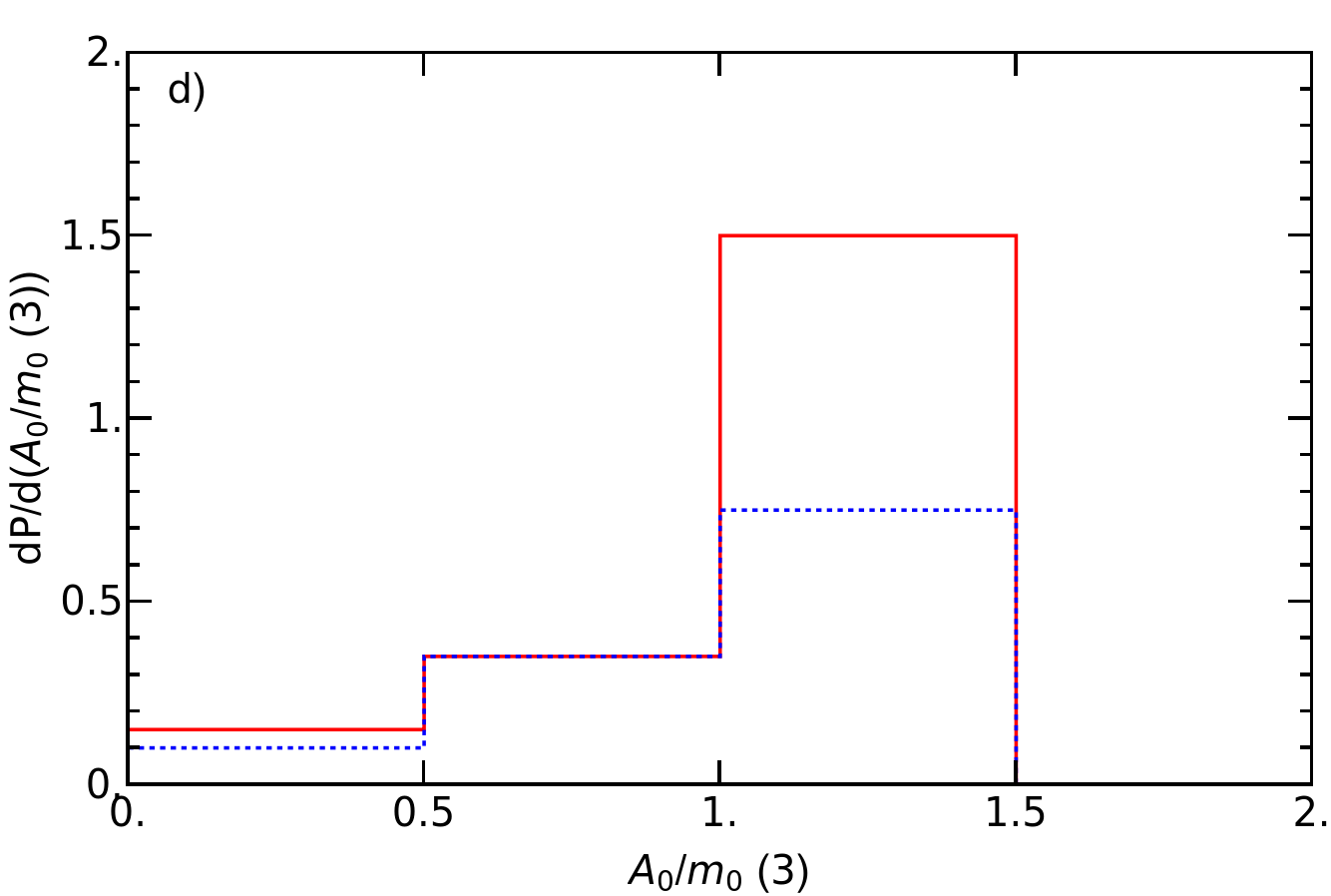}
  \caption{Plot of {\it a}) $dP/dm_{3/2}$, {\it b}) $dP/dm_{0}(1,2)$,
    {\it c}) $dP/dm_{0}(3)$ and {\it d}) $dP/d(A_0/m_0(3))$
    from an $n=1$ landscape scan in the nAMSB model.
The red histogram shows the full probability distribution while the
blue-dashed histogram shows the remaining distribution after
LHC sparticle mass limits are imposed.
    \label{fig:scan1params}}
\end{center}
\end{figure}

In Fig. \ref{fig:scan1params}, we show the probability for selected
nAMSB model input parameters. In frame {\it a}), the distribution
$dP/dm_{3/2}$ rises to a broad peak between $m_{3/2}:100-250$ TeV and cuts off
sharply around 300 TeV. The upper cutoff on $m_{3/2}$ occurs because
as $m_{3/2}\to 300$ TeV, then $m_{\tg}$ is pulled beyond $5-6$ TeV.
In this case, the coupled RGEs pull stop masses so high that
$\Sigma_u^u(\tst_{1,2})$ start contributing too much to the weak scale.
In frame {\it b}), we show the distribution in first/second generation
sfermion soft mass $m_0(1,2)$.
Here, the distribution rises steadily to the scan
upper limit since first/second generation sfermion contributions to the
weak scale $\Sigma_u^u(\tf_{1,2})$ are proportional to the
corresponding fermion Yukawa coupling. This pull to multi-TeV values of
first/second generation squarks and sleptons provides a landscape
amelioration of the SUSY flavor and CP problems\cite{Baer:2019zfl}.
We also show as a black-dashed histogram the results from a special run with
increased upper scan limit of $m_0(1,2)<50$ TeV.
In this case, the distribution peaks at $m_0(1,2)\sim 15-30$ TeV before
getting damped by the anthropic condition that $m_Z^{PU}$ lies within
the ABDS window.
In frame {\it c}), we show the distribution in third generation
soft term $m_0(3)$. In this case, the distribution peaks at $\sim 5$ TeV
albeit with a distribution extending between $2-10$ TeV.
The reason for the upper cutoff is usually that the $\Sigma_u^u(\tst_{1,2})$
contribution to the weak scale becomes too large.
Finally, in frame {\it d}), we show the distribution in the ratio
$A_0/m_0(3)$. This distribution shows the prediction of large bulk $A$-terms
which actually suppress the contributions of $\Sigma_u^u(\tst_{1,2})$
to the weak scale\cite{Baer:2012up}.
But if $A_0$ gets too big, then one is pulled into CCB vacua\cite{Baer:2016lpj}
which fail the anthropic criteria.

\begin{figure}[tbp]
\begin{center}
  \includegraphics[height=0.2\textheight]{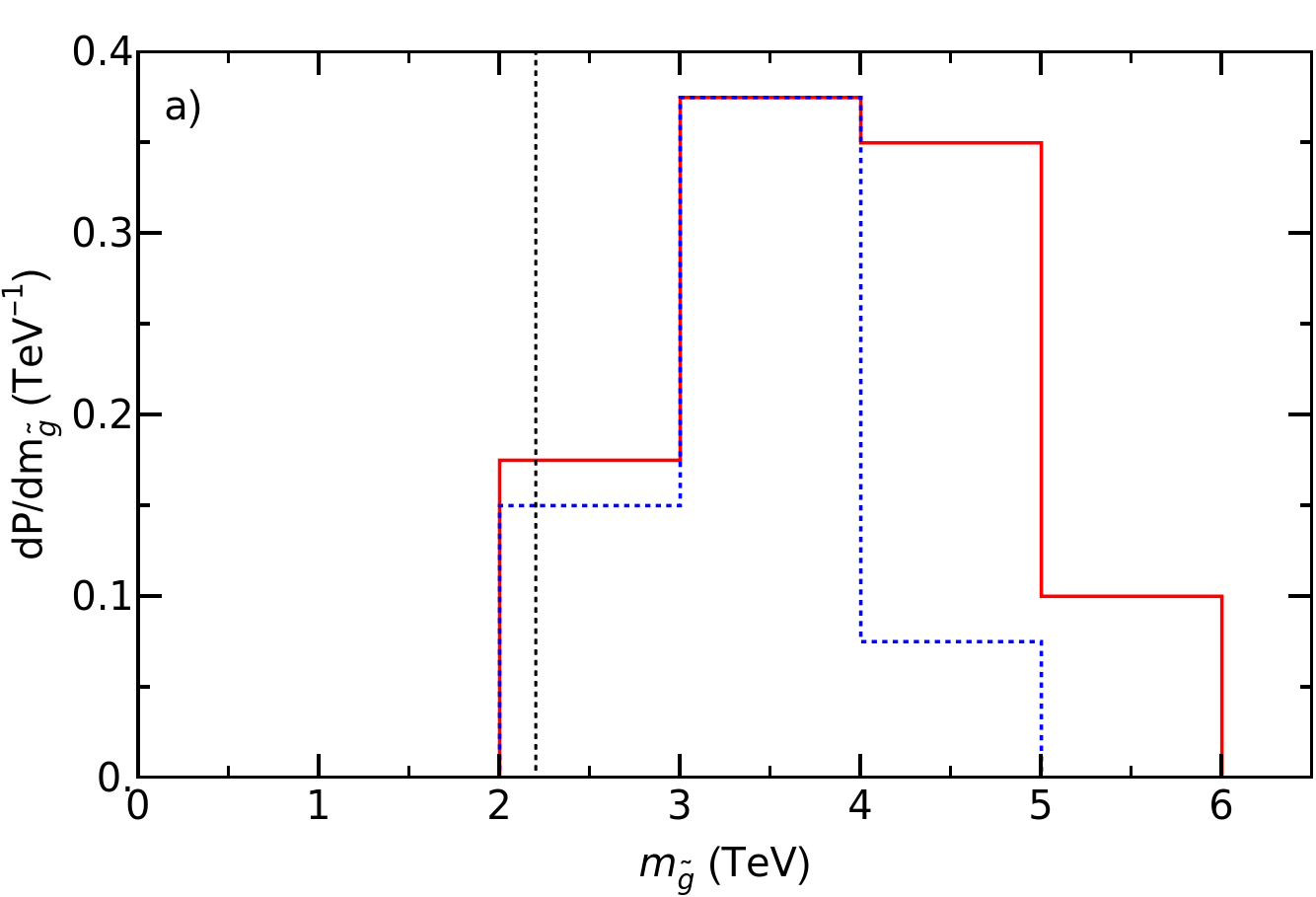}
  \includegraphics[height=0.2\textheight]{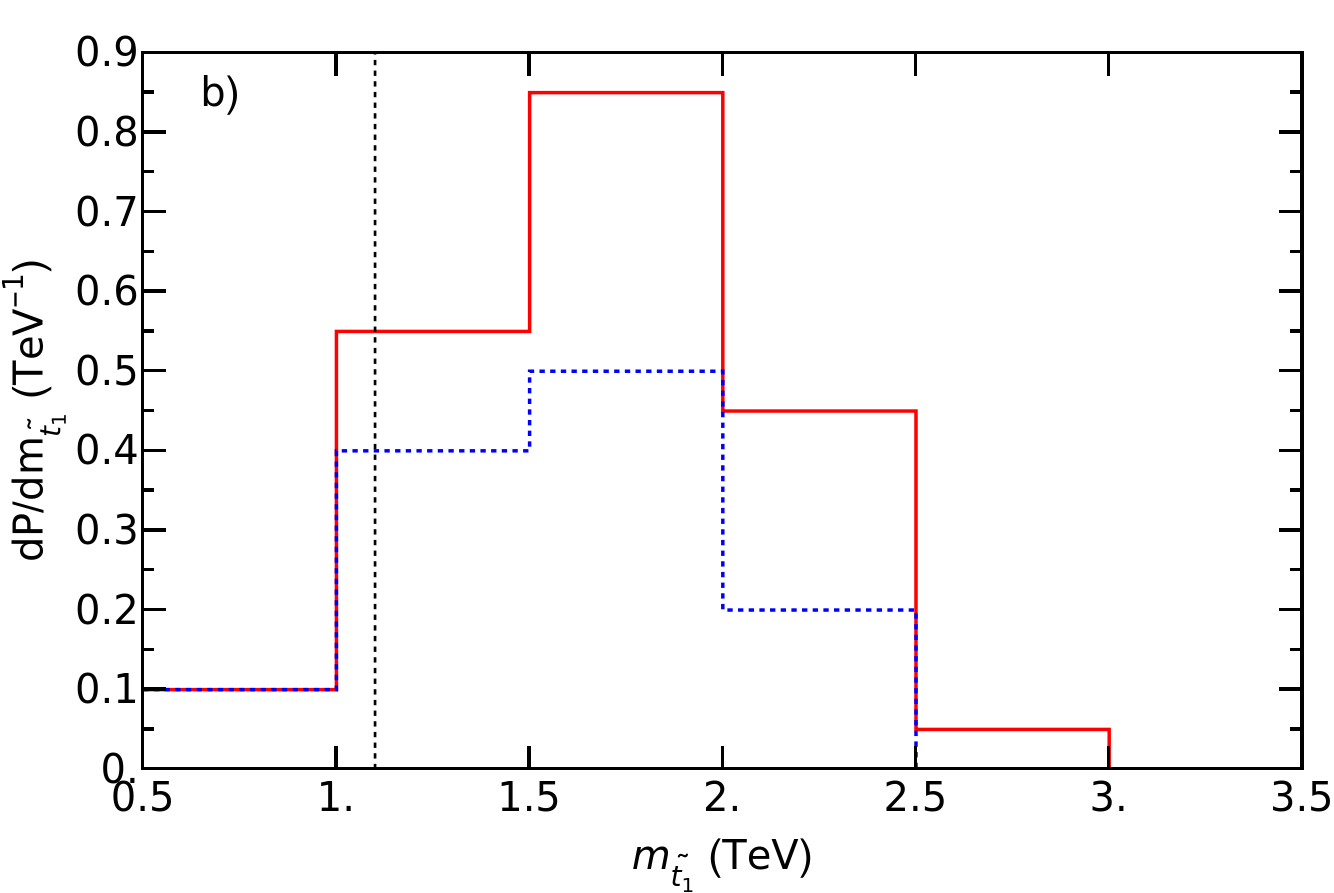}\\
  \includegraphics[height=0.2\textheight]{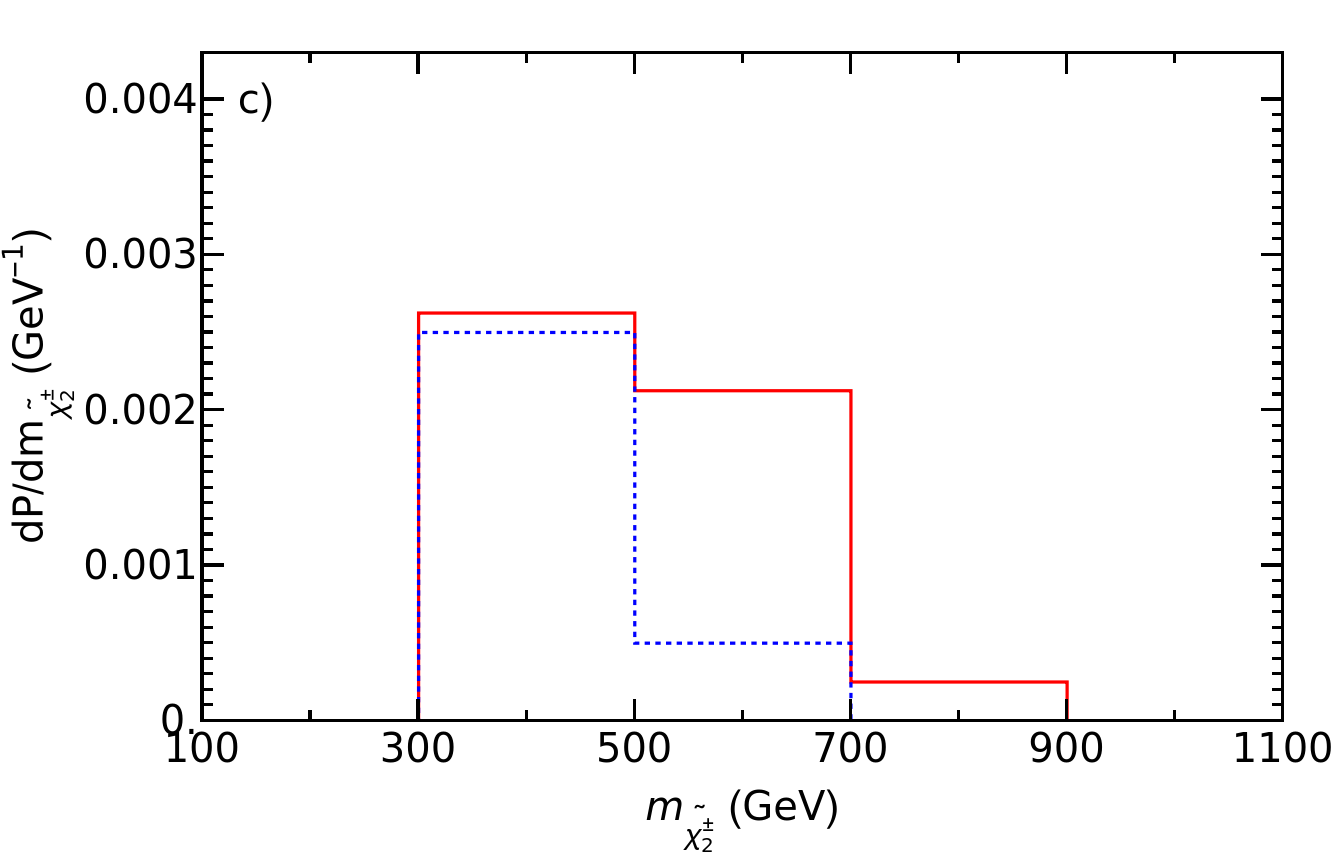}
  \includegraphics[height=0.2\textheight]{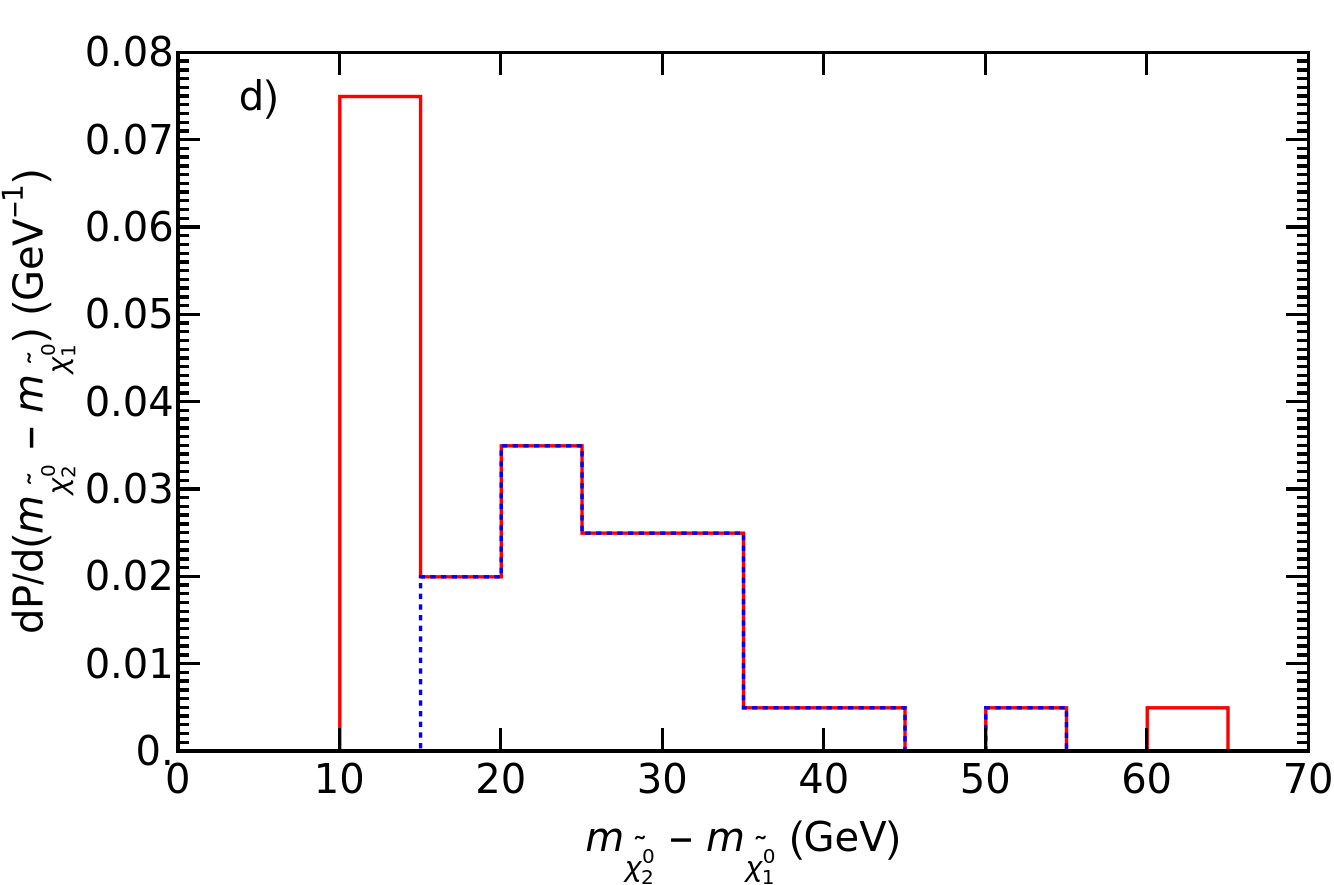}
  \caption{Plot of {\it a}) $dP/dm_{\tg}$, {\it b}) $dP/dm_{\tst_1}$,
    {\it c}) $dP/dm_{\tchi_2^\pm}$ and {\it d}) $dP/d(m_{\tchi_2^0}-m_{\tchi_1^0})$
    from an $n=1$ landscape scan in the nAMSB model.
The red histogram shows the full probability distribution while the
blue-dashed histogram shows the remaining distribution after
LHC sparticle mass limits are imposed.
    \label{fig:scan1masses}}
\end{center}
\end{figure}

In Fig. \ref{fig:scan1masses}, we show the $n=+1$ landscape probability
distribution predictions for various sparticle masses.
In frame {\it a}), we show the distribution in gluino mass $m_{\tg}$.
The distribution begins around $m_{\tg}\sim 2$ TeV and peaks at
$m_{\tg}\sim 3-4.5$ TeV.
This ``stringy natural''\cite{Baer:2019cae} distribution
can explain why it was likely that LHC would not discover weak scale SUSY
via gluino pair production at Run 2, and why gluino pair searches may even
elude HL-LHC searches\cite{Baer:2016wkz}.
The light stop mass distribution is shown in frame {\it b}), and predicts
$m_{\tst_1}\sim 1-2.5$ TeV which is mostly  within range of HL-LHC\cite{Baer:2023uwo}. In frame {\it c}), we show the distribution in $m_{\tchi_2^\pm}$ which is
approximately the wino mass. Here, the bulk of the probability distribution
lies between $M_2\sim 300-700$ GeV, making wino pair production
an inviting target for LHC searches.
In Fig. \ref{fig:scan1masses}{\it d}), we show the distribution in
mass difference of the two lightest neutralinos: $m_{\tchi_2^0}-m_{\tchi_1^0}$.
This mass gap is relevant for the reaction $pp\to\tchi_1^0\tchi_2^0$
where $\tchi_2^0\to f\bar{f}\tchi_1^0$ and thus provides a kinematic
upper bound for the $m(f\bar{f})$ invariant mass. From the distribution, the
mass gap peaks between 10-15 GeV with a tail extending out to 40 GeV
(and even beyond).

\section{$AMSB$ benchmark points and model lines}
\label{sec:BM}

In this Section, we compile three AMSB model benchmark points
using the Isajet 7.91 code\cite{Paige:2003mg} for sparticle and Higgs mass spectra.
The 7.91 version includes several fixes which give better convergence in
the nAMSB model than previous versions.

\subsection{$mAMSB$, $nAMSB0$ and $nAMSB$ benchmark points}

\subsubsection{$mAMSB$ benchmark}

In Table \ref{tab:bm}, we list three AMSB model benchmark points
from three different AMSB models, but with similar underlying parameters
which are convenient for comparison. In Column 2, we list sparticle and
Higgs masses for the usual minimal AMSB model\cite{Gherghetta:1999sw,Feng:1999hg} where universal bulk scalar contributions $m_0^2$ were added to
all AMSB scalar soft masses but no bulk $A_0$ terms were included.
We take $m_{3/2}=125$ TeV and $\tan\beta =10$ with $m_0=5$ TeV.
The $\mu$ term is finetuned to a value $\mu=1719$ GeV to ensure $m_Z=91.2$ GeV,
so the model will be highly finetuned with $\Delta_{EW}=711$ (as listed).
The gluino mass $m_{\tg}=2.73$ TeV so that gluinos are safely beyond LHC
Run 2 search limits which require $m_{\tg}\agt 2.3$ TeV (in simplified models).
The light Higgs mass $m_h=120.3$ GeV: too light compared to its measured value
(and so this BM point is ruled out). The LSP is wino-like with mass $m_{\tchi_1^0}=366$ GeV while $\tchi_2^0$ is binolike and the $\tchi_{3,4}^0$ and $\tchi_2^\pm$
are higgsinolike with mass $\sim \mu$. The top-squark is not very mixed with
$m_{\tst_1}=3.43$ TeV, safely above LHC stop search limits.
With a wino-like LSP, the thermally-produced relic abundance
$\Omega_{\tchi}^{TP}h^2=0.009$, underabundant by a factor $\sim 13$.
Thus, non-thermal wino production mechanisms would need to be active
to fulfill the relic abundance with pure wino dark matter, which
would then be ruled out by indirect WIMP detection experiments,
where winos could annihilate strongly in dwarf galaxies, thus yielding
high energy gamma rays in violation of limits\cite{Baer:2016ucr} from
Fermi-LAT and HESS. Alternatively, a tiny  abundance of wino DM
could be allowed if some other particle such as axions constituted the
bulk of dark matter\cite{Bae:2015rra}.
%
\begin{table}\centering
\begin{tabular}{lccc}
\hline
parameter & mAMSB & nAMSB0 & nAMSB \\
\hline
$m_{3/2}$      & 125000 & 125000 & 125000 \\
$\tan\beta$    & 10 & 10 & 10  \\
$m_0(1,2)$      & 5000 & 10000 & 10000 \\
$m_0(3)$      & 5000 & 5000 & 5000  \\
$A_0$      & 0 & 0 & 6000 \\
\hline
$\mu$          & 1718.8 & 250 & 250   \\
$m_A$          & 5185.3 & 2000 & 2000 \\
\hline
$m_{\tg}$   & 2728.1 & 2803.0 & 2802.4  \\
$m_{\tu_L}$ & 5460.8 & 10211.8 & 10210.9 \\
$m_{\tu_R}$ & 5484.4 & 10289.0 & 10312.1 \\
$m_{\te_R}$ & 4965.1 & 9918.6 & 9889.4 \\
$m_{\tst_1}$& 3428.4 & 3165.3 & 1487.9 \\
$m_{\tst_2}$& 4564.6 & 4253.6 & 3657.9 \\
$m_{\tb_1}$ & 4545.6 & 4238.5 & 3691.4 \\
$m_{\tb_2}$ & 5435.8 & 5239.3 & 5165.7 \\
$m_{\ttau_1}$ & 4918.6 & 4789.1 & 4692.3 \\
$m_{\ttau_2}$ & 4945.1 & 4253.6 & 4978.0 \\
$m_{\tnu_{\tau}}$ & 4945.7 & 4940.9 & 4956.6 \\
$m_{\tchi_2^\pm}$ & 1746.6 & 393.6 & 387.9 \\
$m_{\tchi_1^\pm}$ & 366.2 & 241.2 & 238.1 \\
$m_{\tchi_4^0}$ & 1745.2 & 1176.6 & 1174.2 \\ 
$m_{\tchi_3^0}$ & 1742.4 & 402.4 & 395.0 \\ 
$m_{\tchi_2^0}$ & 1163.0 & 260.9 & 260.4 \\ 
$m_{\tchi_1^0}$ & 366.0 & 229.4 & 226.3 \\ 
$m_h$       & 120.3 & 120.7 & 125.0 \\ 
\hline
$\Omega_{\tchi_1^0}^{TP}h^2$ & 0.009 & 0.01 & 0.01 \\
$BF(b\to s\gamma)\times 10^4$ & 3.1 & 3.2 & 3.3 \\
$BF(B_s\to \mu^+\mu^-)\times 10^9$ & 3.8 & 3.8 & 3.8 \\
$\sigma^{SI}(\tchi_1^0, p)$ (pb) & $7.5\times 10^{-11}$ & $1.8\times 10^{-8}$ & $1.6\times 10^{-8}$ \\
$\sigma^{SD}(\tchi_1^0 p)$ (pb) & $1.7\times 10^{-7}$ & $2.2\times 10^{-4}$ & $2.6\times 10^{-4}$  \\
$\langle\sigma v\rangle |_{v\to 0}$  (cm$^3$/sec)  & $6.1\times 10^{-25}$ & $2.4\times 10^{-25}$& $2.6\times 10^{-25}$ \\
$\Delta_{\rm EW}$ & 711 & 60 & 15.0 \\
\hline
\end{tabular}
\caption{Input parameters and masses in~GeV units
  for the mAMSB, nAMSB0 and nAMSB natural generalized anomaly mediation
  SUSY benchmark points
with $m_t=173.2$ GeV using Isajet 7.91.
}
\label{tab:bm}
\end{table}

\subsubsection{$nAMSB0$ benchmark}

Benchmark point nAMSB0 shows the expected sparticle and Higgs mass spectra
from the generalized AMSB model inspired by DSB where hidden sector singlets
are not allowed. This leads to allowed-- but non-universal--
scalar masses whilst gaugino masses and $A$-terms are suppressed and thus
assume their loop-suppressed AMSB form.
Thus, for nAMSB0 we adopt the parameter space Eq. \ref{eq:nAMSB}
but with $A_0=0$. We adopt a natural value of $\mu =250$ GeV with
$m_A=2$ TeV and also allow for higher first/second generation scalar masses
as expected from the landscape, with $m_0(1,2)=10$ TeV whilst $m_0(3)=5$ TeV
as in the mAMSB benchmark point.

For nAMSB0, the natural value of $\mu =250$ GeV implies light higgsinos
so that while winos are still the lightest gauginos, the higgsinos
are the lightest EWinos, and thus the expected phenomenology
markedly changes from mAMSB. The small value of $\mu$ also makes
the nAMSB0 model much more natural than mAMSB, where $\Delta_{EW}$ has
dropped to 60. The dominant contributions to $\Delta_{EW}$ come now from
$\Sigma_u^u(\tst_{1,2})$. But the model is still somewhat unnatural
since the largest contribution to the RHS of Eq. \ref{eq:mzs} is still
$\sim 500$ GeV, outside the ABDS window\cite{Agrawal:1997gf},
and thus in need of finetuning.
Another problem is the light Higgs mass $m_h=120.7$ GeV.
Both of these issues arise from the rather small AMSB0 value for the
trilinear soft terms.

\subsubsection{$nAMSB$ benchmark}

In the fourth column of Table \ref{tab:bm}, we list the nAMSB benchmark
point which could arise from the sequestered SUSY breaking
scenario of RS\cite{Randall:1998uk},
where in addition to bulk scalar masses, bulk $A$-terms are also expected. 
Here, we use the same parameters as in nAMSB0 except now also allow
$A_0=6$ TeV. The large trilinear soft term leads to large stop mixing
which feeds into the $m_h$ value (which is maximal for stop mixing parameter
$x_t\sim \sqrt{6}m_{\tst}$) so that now the value of $m_h$ is lifted to
125 GeV in accord with LHC measurements. Also, the large positive $A$-term
leads to cancellations in both of $\Sigma_u^u({\tst_1})$ and $\Sigma_u^u({\tst_2})$ leading to increased naturalness where now $\Delta_{EW} =15$.
For the nAMSB0 benchmark, the more-mixed lighter stop mass has dropped to
just $m_{\tst_1}\sim 1.5$ TeV, within striking distance of
HL-LHC\cite{Baer:2023uwo}.

\subsection{Corresponding AMSB model lines}

In this Subsection, we elevate each of the AMSB  benchmark points to
AMSB model lines where we keep the auxiliary parameters 
fixed as before but now allow the fundamental AMSB parameter $m_{3/2}$ to vary.
We compute the AMSB model line spectra using Isasugra.

\begin{figure}[tbp]
\begin{center}
\includegraphics[height=0.4\textheight]{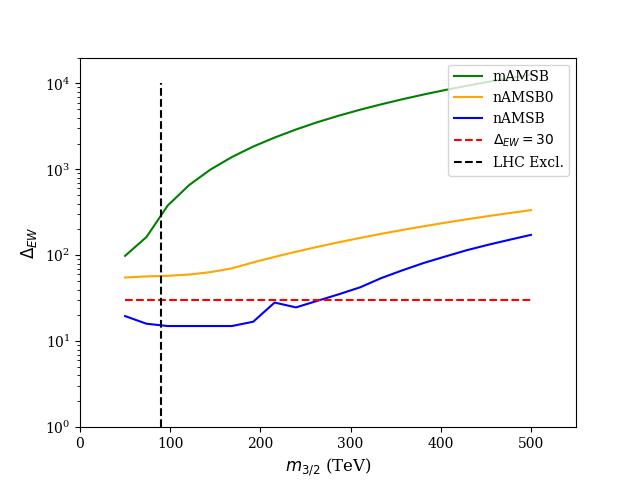}
\caption{Plot of $\Delta_{EW}$ vs. $m_{3/2}$ along the AMSB model lines.
  The region below the dashed line $\Delta_{EW}<30$ is regarded as natural.
  \label{fig:dew}}
\end{center}
\end{figure}

In Fig. \ref{fig:dew}, we first show the naturalness measure
$\Delta_{EW}$ for each model line. For the mAMSB model line, we
see that $\Delta_{EW}$ starts at $\sim 100$ for low $m_{3/2}\sim 50$ TeV,
and then steadily increases to $\Delta_{EW}\sim 10^4$ for
$m_{3/2}\sim 500$ TeV. As for the mAMSB BM point, the dominant
contribution to $\Delta_{EW}$ comes from the (finetuned) $\mu$ parameter.
This model line thus seems highly implausible for all $m_{3/2}$
values based on naturalness. We also show the nAMSB0 model line as the
orange curve. Here, $\Delta_{EW}$ ranges from $50-200$ as $m_{3/2}$
varies over $50-500$ TeV. While more natural than mAMSB, it still
lies outside the ABDS window which is typified by $\Delta_{EW}\alt 30$.
The blue curve shows the nAMSB model line. In this case,
$\Delta_{EW}$ ranges from $\sim 15-150$. The line $\Delta_{EW}=30$ is shown by
the dashed red curve. Here, we see the model line starts becoming unnatural
for $m_{3/2}\agt 265$ TeV.

\begin{figure}[tbp]
\begin{center}
\includegraphics[height=0.4\textheight]{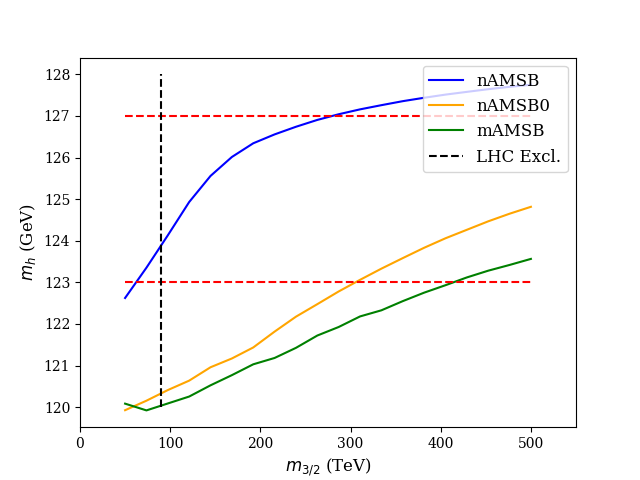}
\caption{Plot of $m_h$ vs. $m_{3/2}$ along the AMSB model lines.
  The light Higgs mass is constrained by LHC measurements to lie between
  the dashed lines, given some theory error on the calculation of $m_h$.
  \label{fig:mh}}
\end{center}
\end{figure}

In Fig. \ref{fig:mh}, we show the computed value of $m_h$ along the three
model lines. The LHC measured window is between $m_h: 123-127$ allowing for a
$\pm 2$ GeV theory error in the computed value of $m_h$.
We see that the mAMSB model line enters the allowed region of $m_h$ only for
$m_{3/2}\agt 400$ TeV while the nAMSB0 model line enters the allowed
$m_h$ range for $m_{3/2}\agt 300$ TeV. Both model lines are highly unnatural
for such large $m_{3/2}$ values. However, the nAMSB model line is within
the $m_h=125\pm 2$ GeV band for $m_{3/2}: 50-280$ TeV, consistent with
its natural allowed range (thanks to the presence of bulk $A_0$ terms).

\begin{figure}[tbp]
\begin{center}
\includegraphics[height=0.4\textheight]{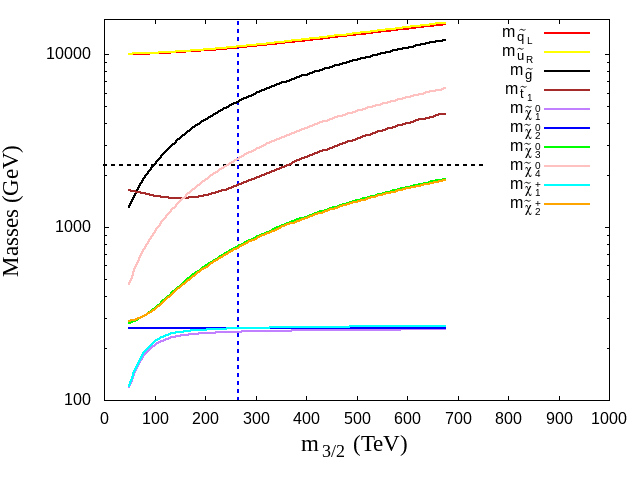}
\caption{Plot of sparticle masses vs. $m_{3/2}$ along the nAMSB model lines.
  \label{fig:masses}}
\end{center}
\end{figure}

In Fig. \ref{fig:masses}, we show various sparticle masses for the
nAMSB model line vs. $m_{3/2}$. The dark and light blue and lavendar lines
show the various higgsino-like EWinos which are typically of order
$m(higgsinos)\sim \mu\sim 250$ GeV. Next heaviest are the wino-like
EWinos $\tchi_3^0$ and $\tchi_2^\pm$, shown as green and orange curves.
These masses vary from $m(winos): 300-2000$ GeV over the range of
$m_{3/2}$ shown, and are, as we shall see, subject to present and future
LHC EWino searches.

The black curve shows the gluino mass $m_{\tg}: 1.2-10$ TeV.
We also show the LHC lower bound $m_{\tg}\sim 2.3$ TeV from gluino pair
searches within the context of simplified models by the black dashed line. 
The LHC simplified model results should apply well in the case of nAMSB
models since the $\tg-\tchi_1^0$ mass gap is aways substantial.
The LHC $pp\to \tg\tg X$ search limits thus provide a lower bound on allowed
nAMSB parameter space with $m_{3/2}\agt 90$ TeV.
The blue dashed line denotes the upper limit on $m_{3/2}$ obtained from
naturalness constraints.
The lighter top squark mass $m_{\tst_1}$ is also shown, and is beyond the
LHC simplified model limit $m_{\tst_1}\agt 1.1$ TeV for all $m_{3/2}$ values.
First/second generation sfermion masses lie around $m_0(1,2)$ value
so in this case would be inaccessible to present and future LHC searches.
By combining lower limits from LHC gluino pair searches with upper bounds
from naturalness, we expect the allowed $m_{3/2}$ values for nAMSB to lie
between $m_{3/2}: 90-265$ TeV.

\section{LHC production cross sections}
\label{sec:prod}

\begin{figure}[tbp]
\begin{center}
\includegraphics[height=0.4\textheight]{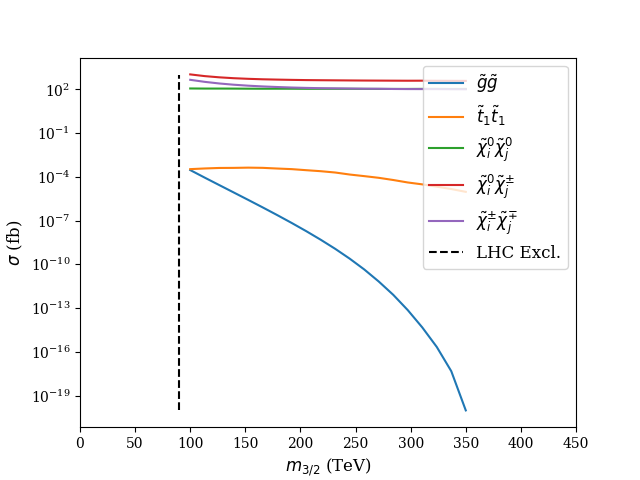}
\caption{Plot of $\sigma (pp\to\tg\tg,\ \tst_1\tst_1^*)$ and EWino
  pair production vs. $m_{3/2}$ along the nAMSB model line.
  \label{fig:sigma}}
\end{center}
\end{figure}

In this Section, we pivot to prospects for LHC searches for SUSY within
the context of the nAMSB model. First, we adopt the computer code
PROSPINO\cite{Beenakker:1996ed} to compute the NLO production cross sections
for various $pp\to SUSY$ reactions, given input from the Isajet
SUSY Les Houches Accord (SLHA) file\cite{Skands:2003cj}.
Our first results are shown in Fig. \ref{fig:sigma} where we show cross
sections for $pp\to \tg\tg$, $\tst_1\tst_1^*$ and (summed) EWino pair production
vs. $m_{3/2}$ along the nAMSB model line.
At the top of the plot,
we see EWino pair production is dominant and relatively flat vs. $m_{3/2}$
since it is dominated by higgsino pair production and $\mu$ is fixed at 250 GeV.
The EWino cross section are divided up into summed $\tchi_i^0\tchi_j^0$,
$\tchi_i^0\tchi_k^\pm$ and $\tchi_k^\pm\tchi_l^\mp$ production,
where $i,j=1-4$ and $k,l=1-2$. The summed EWino pair cross sections are all comparable and of order $\sim 10^2$ fb. The $pp\to\tst_1\tst_1^*$ cross section
is also relatively flat, this time reflecting that $m_{\tst_1}$ hardly changes
with increasing $m_{3/2}$ (from Fig. \ref{fig:masses}). The $pp\to\tg\tg$
cross section is falling rapidly with increasing $m_{3/2}$, reflecting that
the gluino mass is directly proportional to $m_{3/2}$.
From the plot, we thus expect most of the reach of LHC for the
nAMSB model will come from EWino pair production rather than from
gluino or stop pair production.

\begin{figure}[tbp]
\begin{center}
  \includegraphics[height=0.25\textheight]{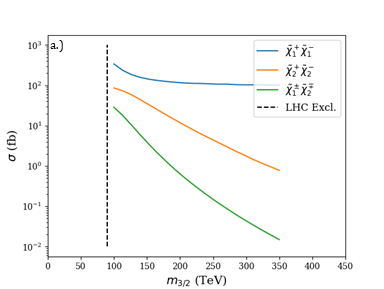}
  \includegraphics[height=0.25\textheight]{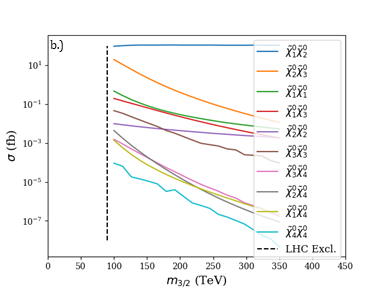}\\
  \includegraphics[height=0.25\textheight]{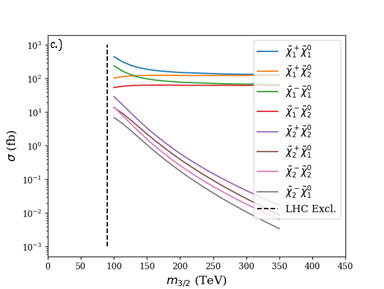}
    \includegraphics[height=0.25\textheight]{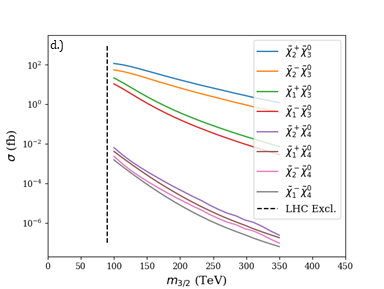}
  \caption{Plot of various EWino pair production cross sections 
    vs. $m_{3/2}$ along the nAMSB model line: {\it a}) chargino pair production, {\it b}) neutralino pair production, {\it c}) chargino-$\tchi_{12}^0$
    pair production and {\it d}) chargino-$\tchi_{3,4}^0$ pair production.
  \label{fig:EWinos}}
\end{center}
\end{figure}

There are many subreactions that contribute to the summed EWino pair
production cross sections. Each subreaction leads to different final states
and thus different SUSY search strategies.
In Fig. \ref{fig:EWinos}{\it a}), we show the several chargino-chargino
pair production reactions vs. $m_{3/2}$.
The upper blue curve denotes $\tchi_1^+\tchi_1^-$ where the light
charginos are mainly higgsino-like (except for some substantial mixing at low $m_{3/2}$ where the wino soft term $M_2\sim \mu$).
Given the small $m_{\tchi_1^+}-m_{\tchi_1^0}$ mass gap, where much of the reaction
energy goes into the invisible LSP mass and energy,
this reaction is likely to be largely invisible at LHC.
The orange curve denotes charged wino pair production:
$\tchi_2^+\tchi_2^-$. Given its modest size and the branching fractions
from Sec. \ref{sec:decay}, it can be very promising for LHC searches.
The third reaction, mixed higgsino-wino $\tchi_1^\pm\tchi_2^\mp$
production, occurs at much lower rates.

In Fig. \ref{fig:EWinos}{\it b}), we show the ten neutralino pair production reactions $\sigma (pp\to \tchi_i^0\tchi_j^0)$. By far, the dominant
neutralino pair production reaction is $pp\to \tchi_1^0\tchi_2^0$.
This reaction takes place dominantly via $s$-channel $Z^*$ exchange
involving the coupling $W_{ij}$ of Eq. 8.101 of Ref. \cite{Baer:2006rs}.
The signs of the neutralino mixing elements add constructively in this
case leading to a large higgsino pair production reaction that leads to
promising LHC signature in the soft opposite-sign dilepton plus jets
plus $\eslt$ channel\cite{Han:2014kaa,Baer:2014kya} (OSDLJMET).
This cross section is flat with increasing
$m_{3/2}$ since $\mu$ is not a soft term and not expected to directly scan in the landscape, but instead is fixed by whatever solution to the SUSY
$\mu$ problem attains\cite{Bae:2019dgg}.
The next largest neutralino pair production cross section is
$pp\to\tchi_2^0\tchi_3^0$: wino-higgsino production, which again has a
constructive sign interference along with large mixing terms.
Other neutralino pair production reactions are subdominant and
typically decreasing with increasing $m_{3/2}$.

In Fig. \ref{fig:EWinos}{\it c}), we show $\tchi_{1,2}^0\tchi_k^\pm$
pair production reactions. The largest, $\tchi_1^0\tchi_1^+$,
may again be largely invisible to LHC searches while the second largest
$\tchi_2^0\tchi_1^+$ can contribute to the OSDLJMET signature mentioned above.
The corresponding reactions with negative charginos are comparable to these
reactions but somewhat suppressed since they occur mainly via $s$-channel
$W^*$ production and LHC is a $pp$ collider which favors positively charged
$W$ bosons.
The remaining higgsino-wino production reactions fall with increasing
$m_{3/2}$ and are subdominant.

In Fig. \ref{fig:EWinos}{\it d}), we show the $\tchi_{3,4}^0\tchi_k^\pm$
production rates. In this case, wino pair production $\tchi_3^0\tchi_2^+$
is dominant but falling as $m_{3/2}$--and hence $M_2$-- increases in value.
The conjugate reaction $\tchi_3^0\tchi_2^-$ reaction is next largest,
followed by $\tchi_3^0\tchi_1^\pm$. The reactions involving bino production
$\tchi_4^0$ are all subdominant and may not be so relevant for LHC
SUSY searches.

\section{Sparticle decay modes}
\label{sec:decay}

In this Section, we wish to comment on some relevant sparticle branching
fractions leading to favorable final state search signatures for LHC.
It is evident from the preceeding Section that EWino pair
production is the dominant sparticle production mechanism at LHC14.
The reaction $pp\to\tchi_1^0\tchi_2^0$ (neutral higgsino pair production)
is dominant, where $\tchi_2^0\to f\bar{f}\tchi_1^0$ and where the $f$ are SM
fermions. For the case of nAMSB, the mass gap $m_{\tchi_2^0}-m_{\tchi_1^0}$
can range up to 50-60 GeV when winos are light, leading to substantial
wino-higgsino mixing for lower values of $m_{3/2}\sim 100$ TeV.
The lucrative leptonic branching fraction $\tchi_2^0\to\ell^+\ell^-\tchi_1^0$
occurs typically at the 2\% level due to competition with other decay modes
such as $\tchi_2^0\to\tchi_1^\pm f\bar{f}^\prime$.

\begin{figure}[tbp]
\begin{center}
  \includegraphics[height=0.3\textheight]{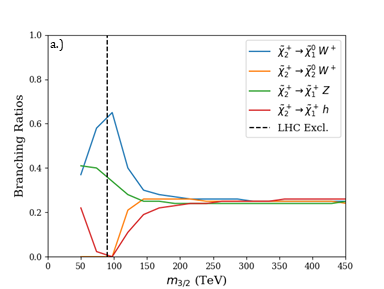}\\
  \includegraphics[height=0.3\textheight]{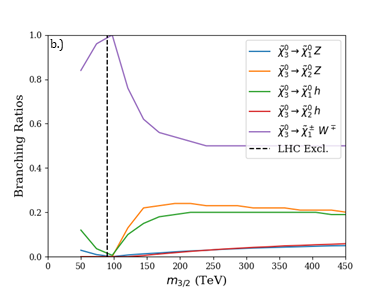}
  \caption{Plot of charged and neutral wino branching fractions
    {\it a}) BF($\tchi_2^+$) and {\it b})
    BF($\tchi_3^0$) versus $m_{3/2}$ along the nAMSB model line.
  \label{fig:winoBFs}}
\end{center}
\end{figure}

The other lucrative production mode from the previous Section is
wino pair production $pp\to\tchi_3^0\tchi_2^\pm$.
To assess the expected final states from this reaction, we plot
in Fig. \ref{fig:winoBFs} the major wino decay branching fractions along
the nAMSB model line. In frame {\it a}), we plot the $BF(\tchi_2^+)$
values vs. $m_{3/2}$ while in frame {\it b}) we plot the
$BF(\tchi_3^0)$ values.
From frame {\it a}), the region with $m_{3/2}\alt 90$ TeV is already
excluded by LHC $\tg\tg$ searches (albeit in the context of simplified models).
Below 90 TeV, there is actually a level-crossing: since $\mu$ is fixed at
250 GeV, a low enough value of $m_{3/2}$ leads to $m(wino)<m(higgsino)$
and an {\it increased} $m_{\tchi_2^+}-m_{\tchi_1^+}$ mass gap
(see Fig. \ref{fig:masses}) so that $\tchi_2^+\to\tchi_1^+ h$ is allowed.
Then, as $m_{3/2}$ increases, the mass gap drops
(due to wino-higgsino degeneracy) and the $\tchi_2^+\to \tchi_1^+h$ mode
becomes kinematically closed. As $m_{3/2}$ increases beyond $\sim 100$ TeV,
then $\tchi_2^+$ becomes wino-like, and the mass gap enlarges so that
the decay $\tchi_2^+\to\tchi_1^+ h$ becomes allowed again. As $m_{3/2}$
increases further, then all four decay modes
$\tchi_2^+\to \tchi_1^0 W^+,\ \tchi_2^0 W^+,\ \tchi_1^+ Z$ and $\tchi_1^+ h$
asymptote to $\sim 25\%$. Thus, we expect the charged wino to decay to
higgsino plus $W$, $Z$ or $h$ in a ratio $\sim 2:1:1$.
Since the higgsinos may be quasi-visible (depending on decay mode and mass gap),
then we get wino decay to $W$, $Z$ or $h$ $+$quasi-visible higgsinos as a
final state.

In Fig. \ref{fig:winoBFs}{\it b}), we show the neutral wino $\tchi_3^0$
branching fractions along the nAMSB model line. At low $m_{3/2}\sim 90$ TeV
near the LHC-excluded region, the neutral winos decay nearly 100\% into
$\tchi_1^\mp W^\pm$. As $m_{3/2}$ increases, the wino-higgsino mass gap
increases, and decays to $\tchi_2^0 Z$ and $\tchi_1^0 h$ are allowed
and can occur at the $\sim 20\%$ level while decays to $\tchi_1^\mp W^\pm$
asymptote to $\sim 50\%$. The remaining branching fraction goes to
mixing-suppressed modes. Thus, for wino pair production, we expect
a final state of $VV+MET$, $Vh+MET$ and $hh+MET$ where $MET$ stands for
missing transverse energy and $V$ stands for the vector bosons $W$ and $Z$.
The $MET$ may not really be entirely
missing since it may include 3-body decay products of the heavier higgsinos.

\begin{figure}[tbp]
\begin{center}
  \includegraphics[height=0.3\textheight]{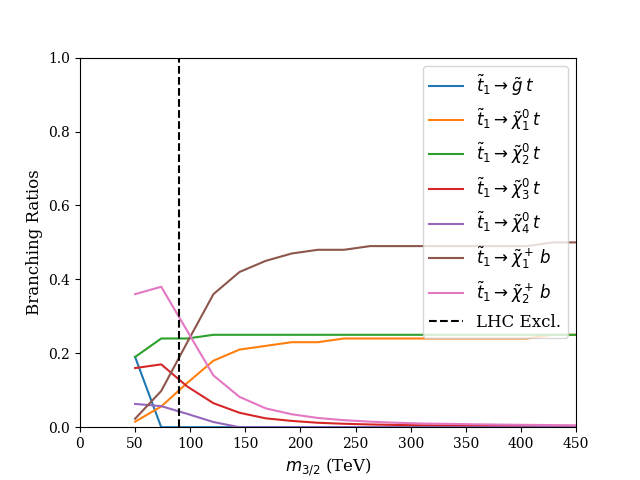}
  \caption{Plot of top squark branching fractions
    BF($\tst_1$) versus $m_{3/2}$ along the nAMSB model line.
  \label{fig:t1BFs}}
\end{center}
\end{figure}

In Fig. \ref{fig:t1BFs}, we plot the light top squark branching fractions
$BF(\tst_1 )$ vs. $m_{3/2}$ along the nAMSB model line. For very low $m_{3/2}$,
$\tst_1\to b\tchi_2^+$ is dominant where the $\tchi_2^\pm$ is mixed
wino-higgsino. But as $m_{3/2}$ increases, $BF(\tst_1\to b\tchi_1^+ )$ becomes
dominant and approaches 50\%, not unlike natural SUSY models with gaugino mass
unification\cite{Baer:2016bwh}. The $\tst_1$ in nAMSB (as in NUHM2 models) is
dominantly $\sim \tst_R$ in spite of large stop mixing soft term $A_t$.
Also, at larger $m_{3/2}$ values, $BF(\tst_1\to t\tchi_{1,2}^0)$
each approach $\sim 25\%$.

\section{LHC excluded regions}
\label{sec:lhc_excluded}

Certain regions of nAMSB model parameter space seem already excluded by
existing LHC13 search limits from Run 2 with $\sim 139$ fb$^{-1}$ of
integrated luminosity.

\subsection{LHC constraint from gluino pair searches}
\label{subsec:gl}
In the case of gluino pair production, for the bulk of
LHC-allowed nAMSB parameter space, we expect $\tg\to t\tst_1^*$ followed
by further $\tst_1$ cascade decays.
The approximate ATLAS and CMS simplified model limits for $\tg\tg$ production
followed by decay to third generation particles should roughly
apply\cite{ATLAS:2022ihe,CMS:2019zmd,CMS:2019ybf},
and these imply
\be
m_{\tg}\agt 2.3\ {\rm TeV} .
\ee
From Fig. \ref{fig:masses}, this implies that $m_{3/2}\agt 90$ TeV.

\subsection{LHC constraint from EWino pair production followed by
  decay to boosted dijets}
  \label{subsec:atlas}

\begin{figure}[tbp]
\begin{center}
\includegraphics[height=0.4\textheight]{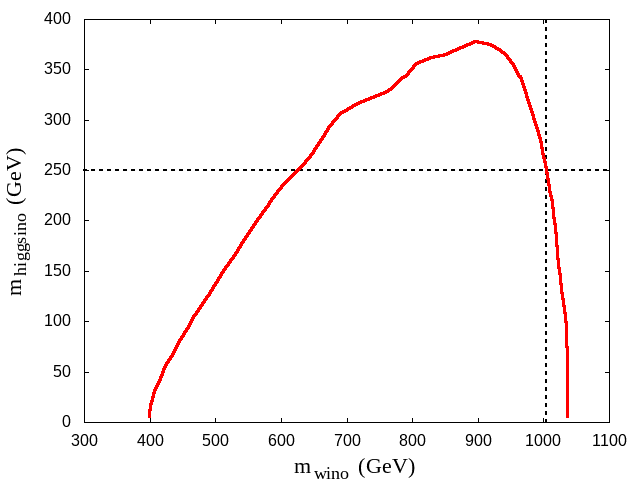}
\caption{Allowed/excluded regions of $m(wino)$ vs. $m(higgsino)$ plane
  from ATLAS analysis of EWino pair production followed by decay to
  $W,Z,h$ with decay to boosted dijets.
  \label{fig:mwinomhiggsino}}
\end{center}
\end{figure}

A recent ATLAS study\cite{ATLAS:2021yqv} reports searching for EWino pair production
followed by two-body decays to $W$, $Z$ or $h$. These heavy SM objects
are assumed to decay hadronically to boosted dijet/fat-jet states
which are then identified.
A similar study by CMS was also made\cite{CMS:2022sfi}, but with smaller
parameter space exclusion regions.
The simplified model limits presented in Fig. 14{\it c}) of
Ref. \cite{ATLAS:2021yqv} should roughly apply to our case for
wino-pair production $pp\to\tchi_2^\pm\tchi_3^0$ as shown in
Fig. \ref{fig:EWinos}{\it d}) followed by decays to vector bosons and
Higgs bosons as shown in Fig. \ref{fig:winoBFs}. The digitized ATLAS
exclusion curve is shown in Fig. \ref{fig:mwinomhiggsino} in the
$m(wino)$ vs. $m(higgsino)$ plane. Our nAMSB model line with
$\mu =250$ GeV is denoted by the horizontal dashed line. From the plot,
we would expect that the range $m(wino):625-1000$ GeV would be ruled
out, corresponding to a range of $m_{3/2}:200-350$ TeV. For model lines
with larger or smaller values of $\mu$, the exclusion region changes
accordingly in Fig. \ref{fig:mwinomhiggsino}.

\subsection{LHC constraints from SModelS/CheckMATE2 analysis}

To test for further limits on nAMSB parameter space,
we employ two recent recasting softwares:
\texttt{SModelS}\cite{Kraml:2013mwa,Ambrogi:2018ujg,Alguero:2021dig} and \texttt{CheckMATE2} (CM2)\cite{Drees:2013wra, Dercks:2016npn} to study
the impact of the current searches on nAMSB parameter space.
\texttt{SModelS} is a popular tool for interpreting simplified-model results
from the LHC. It decomposes Beyond the
Standard Model (BSM) collider signatures presenting a $Z_2$-like symmetry
into Simplified Model Spectrum (SMS) topologies and compares the BSM predictions
for the LHC in a model independent framework with the relevant
experimental constraints.
The main variable for comparison of a BSM theory to the LHC experimental
searches is the $r$-ratio which is defined as the ratio of the
expected $\sigma\times BR$ for a specific final state to the corresponding
upper limit on the $\sigma\times BR\times\epsilon$
(where $\epsilon$ is the acceptance efficiency provided by the
experimental paper).
\texttt{CheckMATE2} is a reinterpretation software for interpreting LHC results for all
BSM models. It is based on recasting the full experimental analyses using events
after full Monte Carlo simulation, hadronization and detector smearing of the
final state objects and implementing the cuts as in the experimental analyses.
It provides the $r$ value defined as the ratio of the expected
number of events from the signal, after implementing all cuts,
to the $95\%$CL  upper limit from the experimental result.
In both cases, for a BSM model to be allowed by current constraints, one
requires $r < 1$.

\begin{figure}[ht]
\begin{center}
\includegraphics[scale=0.5]{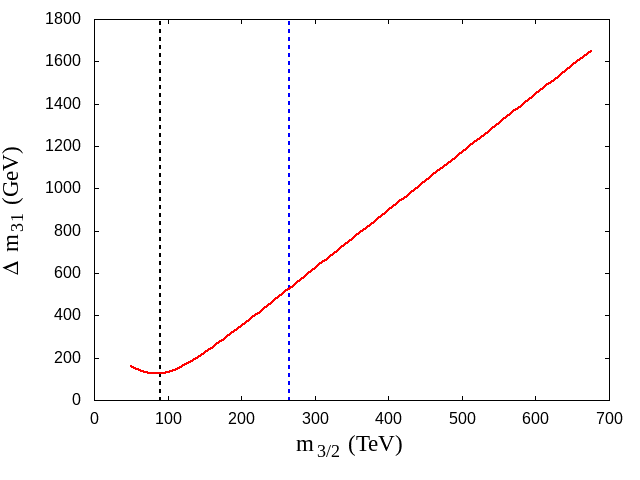} 
\caption{Variation of $\Delta m_{31} = m_{\widetilde{\chi}^0_3} - m_{\widetilde{\chi}^0_1}$ vs. $m_{3/2}$. The region left of the black-dashed curve is excluded by
  LHC13 gluino pair searches and the region to the right of the blue-dashed
line is unnatural with $\Delta_{EW}\agt 30$. }
\label{fig:DelM}
\end{center}
\end{figure}

\begin{figure}[tbp]
\begin{center}
\includegraphics[height=0.4\textheight]{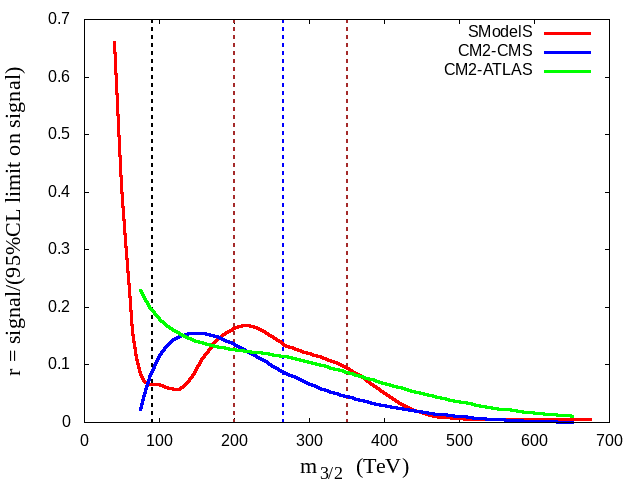}
\caption{Plot of $r$ from \texttt{SModelS} and \texttt{CheckMate2}
  vs. $m_{3/2}$ along the nAMSB model line. 
  \label{fig:r}}
\end{center}
\end{figure}

The wino-higgsino mass gap, quantified by
$\Delta m_{31}=m_{\widetilde{\chi}^0_3} - m_{\widetilde{\chi}^0_1}$, increases with
$m_{3/2}$ as seen in Fig.~\ref{fig:DelM}. 
Fig.~\ref{fig:r} shows the variation of  the highest $r$ value obtained
from \texttt{SModelS} and \texttt{CheckMATE2} for the $\sqrt{s}=$ 13 TeV
results from LHC. The highest $r$-value defined as the ratio of the
signal over the 95$\%$CL upper limit from the signal region is plotted
against $m_{3/2}$. The red dotted lines denote the constraints from the ATLAS search of boosted hadronically decaying bosons + $\eslt$\cite{ATLAS:2021yqv} while the black dotted line denote the bound from the gluino searches implying $m_{3/2}\geq 90$ TeV as discussed in Section \ref{subsec:atlas} and \ref{subsec:gl} respectively.

For the constraints from the \texttt{CheckMATE2} CMS results (blue),
we observe the tightest constraints arise from the multi-lepton (2/3)
$+ \eslt$ searches\cite{CMS:2017moi} for $m_{3/2}\sim 150 $ TeV
with $r$-value $\sim 0.15$ and it falls off on either side of the peak.
This is due to other searches gaining more importance such as searches for
$\geq 4 \ell + \eslt$ for larger mass-gaps between the wino-like
and higgsino-like neutralino.
From the \texttt{CheckMATE2} ATLAS result (green), the $r$-value decreases
with increasing $m_{3/2}$ from $r=0.25$ arising from the hadronic searches
of squarks and gluinos~\cite{ATLAS:2019vcq}.  
%


From \texttt{SModelS} (red), the most stringent constraint occurs at $m_{3/2}<100$ TeV from searches of three leptons 
$ + \eslt$\cite{ATLAS:2021moa}. For $m_{3/2} = 100-250$ TeV range, the most stringent constraints arise from the boosted hadronically decaying diboson $+\eslt$   searches  the multi-lepton searches involving two or three leptons$+\eslt$\cite{CMS:2021edw,ATLAS:2021moa} are the most sensitive searches near the peak at $m_{3/2}\sim225$ TeV.  For higher $m_{3/2} = 250-400$ TeV, the dominant constraints arise from the multi-lepton searches and sub dominant constraints arise from the boosted hadronically decaying dibosons $+\eslt$. As $m_{3/2}$ increases, the multijet $+\eslt$\cite{ATLAS:2017mjy} searches start constraining the parameter space dominantly. However, the $r$-value always remains less than 1: thus, the remaining allowed range of $m_{3/2}\sim 90-200$ TeV appears to be
presently allowed. 

\subsection{LHC-allowed nAMSB parameter space}

\begin{figure}[tbp]
\begin{center}
\includegraphics[height=0.4\textheight]{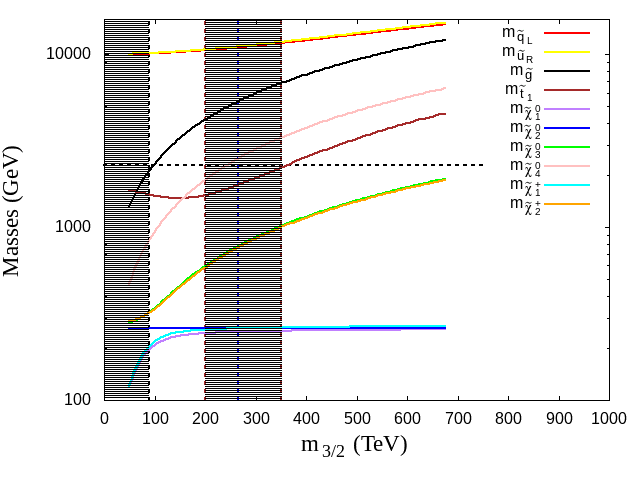}
\caption{Allowed/excluded regions of our nAMSB model-line
  along with various sparticle masses.
    \label{fig:m}}
\end{center}
\end{figure}

Our final allowed nAMSB model line parameter space is shown in
Fig. \ref{fig:m}. The left gray shaded region is excluded by LHC gluino
search limits while the central gray shaded region is excluded by the
ATLAS limits on EWino pair production followed by decay to two boosted
dijet final states. The naturalness limit is denoted by the vertical dashed
line within the central excluded band: the region to the right is unnatural,
and thus highly unlikely (but not impossible) to emerge from the landscape.
The unshaded region extends from $m_{3/2}: 90-200$ TeV and is thus the
presently allowed parameter space.
For convenience, we display again the sparticle masses along our nAMSB
model line.
The remaining SUSY particle spectrum for $m_{3/2}\sim 90-200$ TeV
should provide a target for future
LHC searches seeking to discover or to rule out natural AMSB.

\section{Prospects for nAMSB at Run3 and Hi-Lumi LHC searches}
\label{sec:lhc}

\subsection{LHC higgsino pair production search}

\subsubsection{Soft opposite-sign dilepton, jet$+MET$ search}

In models with light higgsinos, as in natural SUSY, a compelling
LHC search reaction\cite{Baer:2011ec} is $pp\to\tchi_1^0\tchi_2^0$ followed by
$\tchi_2^0\to\ell^+\ell^-\tchi_1^0$, where the dilepton pair is
energetically rather soft since its invariant mass a kinematically
bounded by $m_{\tchi_2^0}-m_{\tchi_1^0}$.
By triggering on hard initial state QCD radiation\cite{Han:2014kaa,Baer:2014kya}, then such
soft dilepton $+\eslt$ events can be searched for at LHC.
Prospects for soft dileptons, jets $+\eslt$ events (soft OSDLJMET) at LHC
have been presented in the higgsino discovery plane\cite{Baer:2020sgm} and in
Ref. \cite{Baer:2021srt} where new angular cuts were proposed to aid in discovery.
Recent search results from CMS\cite{CMS:2021edw} and ATLAS\cite{ATLAS:2019lng}  have been presented.

\begin{figure}[tbp]
\begin{center}
\includegraphics[height=0.4\textheight]{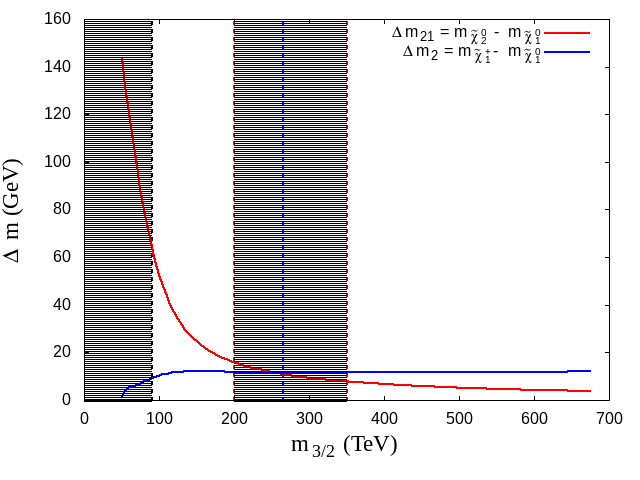}
\caption{Plot of $m_{\tchi_2^0}-m_{\tchi_1^0}$ and $m_{\tchi_1^+}-m_{\tchi_1^0}$
  mass gaps vs. $m_{3/2}$ along the nAMSB model line.
  \label{fig:deltam}}
\end{center}
\end{figure}

The soft OSDJMET signal is a particularly compelling signal for
SUSY in the nAMSB model in light of the large $pp\to\tchi_1^0\tchi_2^0$
cross section from Fig. \ref{fig:EWinos}{\it b}).
A distinguishing feature of the nAMSB model compared to models with
gaugino mass unification or mirage mediation is the relatively larger
$\Delta m_{21}\equiv m_{\tchi_2^0}-m_{\tchi_1^0}$
mass gap ranging from $\sim 15-60$ GeV for nAMSB as shown
in Fig. \ref{fig:deltam} (due to the larger
wino-higgsino mixing from light winos). Current searches from CMS and
ATLAS probe a maximal $\mu$ value of $\sim 200$ GeV for mass gaps
$\Delta m_{21}\sim 10$ GeV. Future ATLAS and CMS probes at HL-LHC
with 3000 fb$^{-1}$ can probe to $\mu\sim 300$ GeV\cite{Canepa:2020ntc}
and the improved angular cuts
may allow HL-LHC to probe as high as $\mu\sim 325$ GeV\cite{Baer:2021srt}.
It should be noted that both ATLAS and CMS seem to have a $2\sigma$ excess
in this channel at present with 139 fb$^{-1}$ of integrated luminosity.
In nAMSB with a larger $m_{\tchi_1^+}-m_{\tchi_1^0}$ mass gap,
soft trilepton plus jet$+\eslt$ signatures should also be available
from $\tchi_1^\pm\tchi_2^0$ production.

\subsection{LHC wino pair production search}

\subsubsection{Boosted hadronic signature}

The rather light winos expected from the allowed parameter-space window
in Fig. \ref{fig:m} provide an inviting target for LHC wino pair production
searches. In the case of $pp\to\tchi_2^\pm\tchi_3^0$ production,
then the relevant signatures occur in the $VV+\eslt$, $Vh+\eslt$
and $hh+\eslt$ channels, where $V=W$ or $Z$. While the strong ATLAS
limits from boosted $V$ or $H$ $\to jj$ already exclude $m_{3/2}:200-350$ TeV,
a search for non-boosted multijets$+\eslt$ may be warranted for
electroweak-produced wino pairs.
These searches may be augmented by searching for the presence of $h\to b\bar{b}$
and $V\to leptons$ in the signal events. New targeted analyses using Run 2
data or forthcoming Run 3 data may even be able to close this allowed window
(or else discover nAMSB SUSY!). Certainly the allowed window in nAMSB
parameter space can be closed by analysis of HL-LHC data.

\subsubsection{Same-sign diboson signature}

The other lucrative search channel for wino pair production followed
by decay to light higgsinos is the same-sign diboson channel
(SSdB)\cite{Baer:2013yha},
where $pp\to\tchi_2^\pm\tchi_3^0$ will be followed by
$\tchi_2^\pm\to W^\pm\tchi_{1,2}^0$ and $\tchi_3^0\to W^\pm\tchi_1^\mp$.
These production and decay modes lead equally to $W^+W^- +\eslt$ and
$W^\pm W^\pm +\eslt$ final states where the former has large SM backgrounds
from $WW$ and $t\bar{t}$ production whilst SM backgrounds for the latter
SSdB signature are far smaller\cite{Baer:2013yha,Baer:2013xua,Baer:2017gzf}.
This relatively jet-free (only jets from initial state QCD radiation)
signature is distinct from the usual same-sign dilepton signature
arising from gluino and squark pair production which should be
accompanied by many hard final state jets.

The reach of HL-LHC for the natural SUSY SSdB signature has been computed
in Ref. \cite{Baer:2017gzf} where peak signal cross sections after cuts
reach the 0.03 fb level compared to total SM backgrounds of $0.005$ fb. 
Whereas the present reach of LHC with 139 fb$^{-1}$ is minimal at
present (for the harder, high luminosity cuts advocated in
Ref. \cite{Baer:2017gzf}), the low wino mass $m(wino)\sim 300-600$ GeV
region should be accessible to LHC Run 3 and HL-LHC data sets
in the 300-3000 fb$^{-1}$ regime. Alternatively, a fresh analysis
by the experimental groups using softer cuts for  the low wino mass
region is clearly warranted.
So far, it seems no dedicated analysis of the SSdB signature
from natural SUSY has been undertaken.

\subsection{LHC stop pair production search: $pp\to\tst_1\tst_1^*$}

Another SUSY search channel for the nAMSB model is via light top
squark pair production $pp\to\tst_1\tst_1^*$ followed by
$\tst_1\to b\tchi_1^+$ at $\sim 50\%$ and $\tst_1\to t\tchi_{1,2}^0$
each at $\sim 25\%$. The reach of HL-LHC for light top-squarks with these
decay modes has been recently evaluated\cite{Baer:2023uwo}.
The $5\sigma$ discovery reach of HL-LHC with 3000 fb$^{-1}$ was found to extend
to $m_{\tst_1}\sim 1.7$ TeV while the 95\% CL reach extended to
$m_{\tst_1}\sim 2$ TeV. These sorts of search limits, performed within
the NUHM2 model, are expected to pertain also to stop pair production
within the nAMSB model.

\section{Summary and conclusions}
\label{sec:conclude}

Supersymmetric models with anomaly-mediated SUSY breaking are well-motivated in
several different SUSY breaking scenarios. In charged SUSY breaking (AMSB0),
gauginos and $A$-terms have suppressed gravity-mediated masses but can
gain dominant AMSB masses whilst scalar masses assume their usual
gravity-mediated form. In the RS AMSB model  with sequestered SUSY breaking,
then gaugino masses, $A$-terms and scalar masses all have the AMSB form,
leading to negative squared slepton masses.
Further bulk scalar mass contributions are required for a viable model.
The phenomenology of mAMSB models is characterized by a wino LSP,
and wino-like WIMP dark matter.
The minimal phenomenological version of these models seems to be triply
ruled out by 1. the difficulty to generate $m_h\sim 125$ GeV unless huge,
unnatural third generation bulk scalar masses are included,
2. the presence of wino-like WIMP dark matter which seems excluded by
direct- and indirect-dark matter detection limits and
3. the large, unnatural value of $\mu$-- and hence large $\Delta_{EW}$--
that such models possess, even for weak-scale soft terms. Rather minor tweaks
to the mAMSB model, already suggested in the original work of
RS\cite{Randall:1998uk}, ameliorate these problems:
non-universal bulk scalar Higgs masses and bulk $A$-terms.
While AMSB0 with non-universal scalar masses still seems ruled out
(due to $A_0\sim 0$ and hence too low $m_h$ values),
the natural AMSB model is both natural
and can accommodate $m_h\sim 125$ GeV. In nAMSB, while the wino is still
the lightest gaugino, the higgsinos are instead the lightest EWinos.
The dark matter issues can be resolved by postulating mixed
axion-higgsino-like WIMP dark matter which is mainly
coposed of axions\cite{Baer:2011hx}.

In this work, we investigated in some detail LHC constraints on natural AMSB
models.
LHC gluino mass limits already require a gravitino mass $m_{3/2}\agt 90$ TeV.
The presence of relatively light winos with mass $m(wino)\sim 300-800$ GeV
implies the model is susceptible to ATLAS/CMS searches for two boosted
dijets $+\eslt$.
Recent ATLAS results seem to rule out $m_{3/2}\sim 200-350$ TeV,
whereas naturalness ($\Delta_{EW}\alt 30$) requires $m_{3/2}\alt 265$ TeV.
The combined constraints leave an open lower mass window of
$m_{3/2}\sim 90-200$ TeV. This lower mass window may soon be excluded (or else
nAMSB may be discovered!) by a combination of
1. soft OS dilepton plus jet$+\eslt$ (OSDLJMET) searches which arise from
higgsino pair production,
2. non-boosted hadronically decaying wino pair production searches and
3. jet-free same-sign diboson searches which are a characteristic signature
of wino pair production followed by wino decay to $W+higgsino$.
Some excess above SM background in the OSDLJMET channel already seems to be
present in both ATLAS and CMS data\cite{ATLAS:2019lng,CMS:2021edw}.

\section*{Acknowledgments}

This material is based upon work supported by the U.S. Department of Energy, 
Office of Science, Office of High Energy Physics under
Award Number DE-SC-0009956.
VB gratefully acknowledges support from the William F. Vilas Estate.


\bibliography{namsb}
\bibliographystyle{elsarticle-num}

\end{document}